\documentclass[12pt]{article}
\usepackage{amsmath}
\usepackage{graphicx}
\usepackage{natbib}
\usepackage{url} % not crucial - just used below for the URL 

\RequirePackage{amsthm,amsmath,amsfonts,amssymb}
\usepackage{enumerate}
\usepackage{natbib}
\usepackage{url} 
\usepackage{float}
\usepackage{graphicx,psfrag,epsf}
\usepackage{enumerate}
\usepackage{booktabs}
\usepackage{framed}  
\usepackage{caption}
\usepackage{pgfplots}
\usepgfplotslibrary{dateplot}
\usetikzlibrary{snakes}
\usepackage{rotating}
\usepackage{multirow, booktabs}
\usepackage{subfig}
\usepackage[T1]{fontenc}
\usepackage[utf8]{inputenc}
\usepackage{xcolor}
\usepackage{tikz}
\usetikzlibrary{positioning, shapes.geometric}
\newcommand{\semaphore}[2]{
\tikz[node distance=3mm,baseline]
    {
    \node (s1) [semicircle, fill=#1, minimum size=3mm] {};
    \node (s1) [semicircle, fill=#2, minimum size=3mm, rotate=180, below=of s1] {};
    }
    }
\colorlet{shadecolor}{gray!50}
\DeclareMathOperator*{\argmax}{argmax} % thin space, limits underneath in displays

\newtheorem{Definition}{Definition}

\newtheorem{Example}{Example}
\newtheorem{Remark}{Remark}

\begin{document}

\def\spacingset#1{\renewcommand{\baselinestretch}%
{#1}\small\normalsize} \spacingset{1}

%%%%%%%%%%%%%%%%%%%%%%%%%%%%%%%%%%%%%%%%%%%%%%%%%%%%%%%%%%%%%%%%%%%%%%%%%%%%%%

\begin{center}
    \Large \bf Sensitivity Analysis for Binary Outcome Misclassification in Randomization Tests via Integer Programming
\end{center}

%Analyzing Randomized Experiments Subject to Outcome Misclassification via Integer Programming

\begin{center}
  \large  $\text{Siyu Heng}^{*}$ and $\text{Pamela A. Shaw}^{\dagger}$
\end{center}

\begin{center}
   \large \textit{$^{*}$ Department of Biostatistics, School of Global Public Health, New York University (Email: siyuheng@nyu.edu)}
\end{center}
\begin{center}
   \large \textit{$^{\dagger}$ Biostatistics Unit, Kaiser Permanente Washington Health Research Institute (Email: Pamela.A.Shaw@kp.org)}
\end{center}

\bigskip
\begin{abstract}
Conducting a randomization test is a common method for testing causal null hypotheses in randomized experiments. The popularity of randomization tests is largely because their statistical validity only depends on the randomization design, and no distributional or modeling assumption on the outcome variable is needed. However, randomization tests may still suffer from other sources of bias, among which outcome misclassification is a significant one. We propose a model-free and finite-population sensitivity analysis approach for binary outcome misclassification in randomization tests. A central quantity in our framework is ``warning accuracy," defined as the threshold such that a randomization test result based on the measured outcomes may differ from that based on the true outcomes if the outcome measurement accuracy did not surpass that threshold. We show how learning the warning accuracy and related concepts can amplify analyses of randomization tests subject to outcome misclassification without adding additional assumptions. We show that the warning accuracy can be computed efficiently for large data sets by adaptively reformulating a large-scale integer program with respect to the randomization design. We apply the proposed approach to the Prostate Cancer Prevention Trial (PCPT). We also developed an open-source \textsf{R} package for implementation of our approach.
\end{abstract}

\noindent%
\vfill

\spacingset{1.75} % DON'T change the spacing!

%%%%%%%%%%%%%%%%%%%%%%%%%%%%%%%%%%%%%%%%%%%%%%%%%%%%%%%%%%%%%

\section{Introduction}

\vspace{-0.2 cm}

\subsection{Outcome misclassification in randomization tests}\label{subsec: introduction} 

For testing causal null hypotheses in randomized experiments (trials), using a randomization test (also known as design-based or finite-population hypothesis testing) is one of the most widely used strategies. Compared with super-population (sampling-based) hypothesis testing, of which theoretical justifications typically require the assumption of independent and identically distributed (i.i.d.) sampling of study subjects from some super-population, randomization tests have two significant characteristics. First, the statistical validity of randomization tests only relies on the randomization of treatment assignments, which can be guaranteed by study design such as randomized experiments, and no distributional assumption (such as the i.i.d. sampling assumption) on the outcome variable is needed. Second, through conditioning on the fixed sequence of potential outcomes, randomization tests focus explicitly on the study subjects in hand instead of some hypothetical super-population, which makes our hypothesis testing results more relevant for the study subjects in hand. For a detailed discussion on the advantages and limitations of randomization tests, see \cite{rosenbaum2002observational}, \cite{feng2014randomization}, \cite{imbens2015causal}, \cite{athey2017econometrics}, \cite{cohen2022gaussian}, \cite{li2023randomization}, and \cite{zhang2023randomization}, among many others. 

Although immune to confounding bias or violations of sampling assumptions on outcomes, randomization tests can still suffer from other sources of bias, among which measurement error in the outcome (\citealp{arnold2016negative} \citealp{hernan2020causal}), referred to as ``binary outcome misclassification" when the outcome is binary (\citealp{carroll2006measurement, buonaccorsi2010measurement}), is a significant one. Outcome misclassification commonly exists in randomized experiments. For example, common binary outcomes in randomized clinical trials are cancer diagnosis (cancerous versus non-cancerous), blood test results (normal versus abnormal), and antibody test results (positive versus negative), among many others. These binary outcomes can be misclassified due to the following factors: technical limitations of the diagnosis methods (\citealp{wittram2004ct}), physician-side factors such as misinterpretations of the clinical data or patient's symptoms (\citealp{walsh1995clinical}), and patient-side factors such as patients' abnormal activities before the diagnosis and anxiety during the diagnosis (\citealp{ogedegbe2008misdiagnosis}). Another major type of outcome misclassification is reporting bias in self-reported binary outcomes, especially those concerning sensitive topics such as unhappiness with the treatment, sexuality questions, and mental health issues (\citealp{catania1986questionnaire, knauper1994diagnosing, wood2008empirical}). 

In general, there are at least two ways that outcome misclassification, in particular, differential outcome misclassification (i.e., the proportions of outcome misclassification are different between the treated and control groups), can severely distort a causal conclusion drawn from a randomization test. First, in randomized trials without blinding or with imperfect blinding, systematically larger effects on subjective outcomes (either subject-reported or investigator-assessed) can occur due to differential outcome misclassification from knowledge of treatment status, as found by a large systematic review of clinical trials (\citealp{wood2008empirical}). Such exaggerated or misleading reporting of health improvements documented in previous non-blinded or imperfectly blinded trials can be due to either placebo effects (i.e., treated subjects tend to perceive or overstate health improvements due to psychological factors) or courtesy bias (i.e., treated subjects tend to not fully state their unhappiness with the treatment as an attempt to be polite toward the investigators). However, perfect blinding is not practical in a wide range of randomized trials, and bias due to outcome misclassification may be inevitable for the downstream randomization tests in such studies (\citealp{arnold2016negative}). Second, in some studies, the treatment can improve or interfere with the detection of the outcome. For example, some clinical trials have found that the treatments they investigated can change the volume of the study organ or the pattern of the study tumor, so diseases may be easier or harder to be detected among treated subjects than control subjects in such cases (\citealp{lucia2007finasteride, redman2008finasteride}). Therefore, even if blinding was effectively carried out in a randomized trial, potential bias due to differential outcome misclassification may still exist and may bias a downstream randomization test (e.g., false rejection of some causal null hypothesis).

\subsubsection{Motivating example: a puzzle from the Prostate Cancer Prevention Trial}\label{subsubsec: PCPT}

Prostate cancer is one of the most common cancers in men. According to \citet{siegel2020colorectal}, prostate cancer is estimated to account for 21\% of the new cancer cases diagnosed in men in the United States in 2020. Since the development of prostate cancer is a long-term process, many studies have focused on the prevention of prostate cancer. Among these studies, the Prostate Cancer Prevention Trial (PCPT) (\citealp{thompson2003influence}) is especially influential because it is ``the first study to show that a drug can reduce a man's chances of developing prostate cancer" (\citealp{nih}). In the PCPT, within each of the 221 study sites, men 55 years of age or older with no evidence of prostate cancer are randomly assigned to finasteride (5 mg per day) or placebo. The primary outcome is whether the participant is diagnosed with prostate cancer during the seven-year follow-up period (including an end-of-study biopsy planned for those who had not been given a diagnosis of prostate cancer during the study). Of the 9060 participants included in the final analysis, 803 of the 4368 ($\approx$ 18.4\%) in the finasteride group and 1147 of the 4692 ($\approx$ 24.4\%) in the placebo group were diagnosed with prostate cancer. Applying the randomization-based Mantel-Haenszel test (one of the most commonly used randomization tests for binary outcomes) to the 221 study sites (strata), the two-sided $p$-value under Fisher's sharp null of no effect is $4.66 \times 10^{-13}$. According to the two-sided 0.05 significance level prespecified in the design stage of the PCPT (\citealp{feigl1995design}), a prevention effect of finasteride on prostate cancer was detected (\citealp{thompson2003influence}).

The PCPT also identified a controversial and seemingly contradictive phenomenon: taking finasteride may increase the risk of high-grade prostate cancer (i.e., tumor with Gleason score $\geq 7$). Specifically, of the 9037 men (out of the total 9060 men in the final analysis) with graded prostate cancer or without prostate cancer, 280 of the 4358 ($\approx 6.4\%$) in the finasteride group and 237 of the 4679 ($\approx 5.1\%$) in the placebo group were diagnosed with high-grade prostate cancer. Applying the Mantel-Haenszel test, the two-sided $p$-value under Fisher's sharp null of no effect is $6.79 \times 10^{-3}$, also statistically significant at the prespecified 0.05 significance level.

Can finasteride prevent prostate cancer but also promote high-grade prostate cancer? This seeming contradiction puzzled many researchers when the results were first published in \textit{New England Journal of Medicine} in 2003 (\citealp{thompson2003influence}). Several follow-up studies have pointed out that this puzzling result could potentially be due to bias caused by misclassification of the prostate cancer status or severity (e.g., \citealp{lucia2007finasteride, redman2008finasteride, shepherd2008does}). Therefore, a natural question is: can we develop a model-free and finite-population sensitivity analysis approach to help explain this puzzle from the perspective of outcome misclassification?

\subsection{The gap in the existing literature and our contributions}\label{subsec: contributions}

Although outcome misclassification often exists in randomized experiments (e.g., randomized clinical trials such as the aforementioned PCPT) and can severely distort the downstream randomization tests, it was often not carefully assessed or addressed in practical research. Although there are many existing approaches (including sensitivity analysis approaches) for randomized experiments with outcome misclassification (e.g., \citealp{magder1997logistic}, \citealp{carroll2006measurement}, \citealp{shepherd2008does}, \citealp{gilbert2016misclassification}, \citealp{shu2019weighted}, \citealp{beesley2020statistical}, \citealp{hubbard2020reducing}, among many others), to our knowledge, there is no established framework for handling outcome misclassification in randomization testing approach to randomized experiments. This is largely due to the challenge that randomization tests do not assume any outcome distributions or models and, therefore, cannot allow researchers to apply the existing model-based (super-population) methods for outcome misclassification without adding strong additional assumptions.

To fill this gap, we develop a model-free and finite-population sensitivity analysis approach for randomization tests subject to outcome misclassification via large-scale integer programming. Our framework provides a unified approach to shed light on the following three common questions in randomization tests subject to outcome misclassification: 1) \textbf{Q1:} How sensitive is a causal conclusion (e.g., rejecting the null effect or not) reported by a randomization test to outcome misclassification? 2) \textbf{Q2:} Is this causal conclusion more sensitive to false positives versus false negatives among the treated versus control subjects? 3) \textbf{Q3:} For a validation substudy on outcome misclassification, how do we choose which subjects' outcomes to validate to make the validation substudy more efficient?

A central concept of our framework is called ``warning accuracy," defined as the threshold such that the causal conclusion (e.g., rejecting some causal null hypothesis or not) based on the measured outcomes may differ from that based on the true outcomes if the finite-population accuracy of the measured outcomes did not surpass that threshold. We show how computing the warning accuracy and related quantities can help answer the aforementioned three questions $\textbf{Q1--3}$ without adding additional assumptions to a randomization test. Therefore, the proposed sensitivity analysis is a free lunch added to the conventional analysis of a randomization test subject to outcome misclassification. To handle the computational challenge encountered in computing warning accuracy for large-scale randomized experiments, called the ``curse of symmetry," we propose a computation strategy to adaptively reformulate a corresponding integer program with respect to the randomization design. Our computation strategy leverages some intrinsic characteristics of various randomization designs and recent advances in erasing symmetry in integer programming (\citealp{fogarty2016discrete, fogarty2017randomization}), which can be of independent interest. Our framework covers both Fisher's sharp null and Neyman's weak null and universally works for several commonly used randomization designs. Our framework can also be applied to matched or stratified observational studies that adopt randomization tests. As a real-data application, we use our sensitivity analysis approach incorporated with related expert knowledge on prostate cancer diagnosis to provide evidence on whether the well-known puzzle concerning the contradictive effects of finasteride detected by the PCPT is due to actual effects or potential outcome misclassification bias. We also developed an open-source \textsf{R} package \textsf{RIOM} (i.e., \textbf{R}andomization \textbf{I}nference with \textbf{O}utcome \textbf{M}isclassification) for implementing our methods and for reproducible research, which is publicly available on \textsf{GitHub} (\url{https://github.com/siyuheng/RIOM}).

Model-based approaches (e.g., model-based average-case sensitivity analyses or bias correction approaches) in the previous literature on outcome misclassification can still be very useful, especially when we want to leverage side information and domain knowledge to study questions beyond the scope of questions \textbf{Q1}-\textbf{Q3}. As long as the models and assumptions are appropriately imposed, researchers can still perform further model-based analyses after adopting our model-free worst-case sensitivity analysis framework. The central spirit of our framework is: Before imposing any additional assumptions to a randomization test to conduct a model-assisted analysis for outcome misclassification, which incurs the risk of assumption violations or model misspecification, what useful information concerning outcome misclassification are we already able to learn from the experimental data?

\vspace{-0.5 cm}

\section{Review}
\vspace{-0.2 cm}

\subsection{Randomization tests with binary outcomes}

Suppose that there are $I\geq 1$ strata, and there are $n_{i}$ subjects in stratum $i$, $i=1,\dots,I$, with $N=\sum_{i=1}^{I}n_{i}$ subjects in total. Let $Z_{ij}$ be the treatment indicator of subject $j$ in stratum $i$: $Z_{ij}=1$ if subject $j$ in stratum $i$ received treatment and $Z_{ij}=0$ otherwise. Suppose that in stratum $i$ a fixed number $m_{i}$ of subjects are designed to receive the treatment, and $n_{i}-m_{i}$ subjects receive control, i.e., $\sum_{j=1}^{n_{i}}Z_{ij}=m_{i}$ where $m_{i}\in \{ 1,\dots, n_{i}-1\}$. Let $\mathbf{Z}=(Z_{11},\dots, Z_{In_{I}})\in \{0,1\}^{N}$ denote the treatment indicator vector and $\mathcal{Z}=\{ \mathbf{Z}\in \{0, 1\}^{N}:\sum_{j=1}^{n_{i}}Z_{ij}=m_{i}, i=1,\dots, I\}$ denote the collection of all possible treatment assignments. In a randomized experiment, the treatment assignments are random within each stratum, i.e.,
\begin{equation}\label{eqn: randomization assumption}
    P(\mathbf{Z}=\mathbf{z}\mid \mathcal{Z})=\prod_{i=1}^{I}{n_{i}\choose m_{i}}^{-1}, \quad \forall \mathbf{z}\in \mathcal{Z}.
\end{equation}
The randomization assumption (\ref{eqn: randomization assumption}) is also widely considered in design-based observational studies after adjusting for confounders using matching or stratification (\citealp{rosenbaum2002observational, rosenbaum2020design, zubizarreta2014matching}). The above setup covers a wide range of randomized experiments and observational studies. When $I=1$, the study is called a completely randomized experiment. For $I\geq 2$, the study is a general stratified/blocked randomized experiment or observational study. When $n_{i}=2$ for all $i$, the study is a paired randomized experiment or pair-matched observational study. When $m_{i}=1$ and $n_{i}=n\geq 3$ for all $i$, the study is a randomized experiment or observational study with multiple controls. When $m_{i}=1$ for all $i$ while $n_{i}$ may vary with $i$, the study is a randomized experiment or observational study with variable controls. When $\min\{m_{i}, n_{i}-m_{i}\}=1$ for all $i$, the study is a finely stratified experiment (\citealp{fogarty2018mitigating}) or an observational study with full matching (\citealp{hansen2004full}). See \citet{imbens2015causal} and \citet{rosenbaum2002observational, rosenbaum2020design} for detailed introductions to various study designs in randomized experiments and design-based observational studies. 

\begin{Remark}
    Following the arguments in Chapter 3 of \citet{rosenbaum2002observational}, equation (\ref{eqn: randomization assumption}) also holds in matched or stratified observational studies assuming (i) exact matching/stratification on measured confounders, (ii) no unmeasured confounding, and (iii) independence across matched sets or stratum. Then, under the above assumptions, randomization tests based on equation (\ref{eqn: randomization assumption}) can be similarly conducted in matched/stratified observational studies as in randomized experiments (\citealp{rosenbaum2002observational, zubizarreta2014matching, fogarty2016discrete}). In some studies, however, the randomization assumption (i.e., equation (\ref{eqn: randomization assumption})) can be violated due to either inexact matching/stratification on measured confounders or the existence of unmeasured confounders. Therefore, researchers often first conduct a randomization test using equation (\ref{eqn: randomization assumption}) as a working assumption, and then conduct a sensitivity analysis to assess sensitivity of the aforementioned randomization test to violation of equation (\ref{eqn: randomization assumption}) (\citealp{rosenbaum2002observational, zubizarreta2014matching, fogarty2017randomization, zhao2019sensitivityvalue}). 
\end{Remark}

Let $Y_{ij} \in \{0,1\}$ be the true outcome of subject $j$ in stratum $i$ and $\mathbf{Y}=(Y_{11}, \dots, Y_{In_{I}})$. Following the potential outcomes framework (\citealp{neyman1923application, rubin1974estimating}), let $Y_{ij}(1)$ and $Y_{ij}(0)$ be the potential true outcome under treatment and that under control of subject $j$ in stratum $i$ respectively, therefore $Y_{ij}=Y_{ij}(1)Z_{ij}+Y_{ij}(0)(1-Z_{ij})$. In a randomization test for randomized experiments or observational studies, potential outcomes are \textit{fixed values}, and the only probability distribution that enters a randomization test is the randomization condition (\ref{eqn: randomization assumption}) (\citealp{rosenbaum2002observational, imbens2015causal}). Fisher's sharp null of no effect on any subject (\citealp{fisher1937design}) asserts that $H_{0}^{\text{sharp}}: Y_{ij}(1)=Y_{ij}(0), \forall i, j$. The commonly used randomization test for testing $H_{0}^{\text{sharp}}$ with binary outcomes are Fisher's exact test (when $I=1$) and the randomization-based Mantel-Haenszel test (when $I\geq 1$), of which the test statistic is defined as $T_{\text{M-H}}(\mathbf{Z}, \mathbf{Y})=\sum_{i=1}^{I}\sum_{j=1}^{n_{i}}Z_{ij}Y_{ij}$ (\citealp{mantel1959statistical}). The Mantel-Haenszel test reduces to Fisher's exact test when $I=1$ and reduces to McNemar's test when $n_{i}=2$ for all $i$ (\citealp{cox2018analysis}). Researchers commonly use the following finite-population central limit theorem (\citealp{rosenbaum2002observational, li2017general}) to test $H_{0}^{\text{sharp}}$: $\frac{T_{\text{M-H}}-E(T_{\text{M-H}})}{\sqrt{\text{Var}(T_{\text{M-H}}) }}\xrightarrow{\mathcal{L}} N(0,1)$ under $H_{0}^{\text{sharp}}$, where $E(T_{\text{M-H}})=\sum_{i=1}^{I}(\frac{m_{i}}{n_{i}}\sum_{j=1}^{n_{i}}Y_{ij})$ and $\text{Var}(T_{\text{M-H}})=\sum_{i=1}^{I}\frac{m_{i}(\sum_{j=1}^{n_{i}}Y_{ij})(n_{i}-\sum_{j=1}^{n_{i}}Y_{ij})(n_{i}-m_{i})}{n_{i}^{2}(n_{i}-1)}$. In a two-sided level-$\alpha$ testing procedure, researchers reject $H_{0}^{\text{sharp}}$ if and only if $\frac{\{T_{\text{M-H}}-E(T_{\text{M-H}})\}^{2}}{\text{Var}(T_{\text{M-H}})}>\chi^{2}_{1, 1-\alpha}$, where $\chi^{2}_{1, 1-\alpha}$ is $1-\alpha$ quantile of chi-squared distribution with one degree of freedom.

In addition to Fisher's sharp null $H_{0}^{\text{sharp}}$, another extensively considered null hypothesis is Neyman's weak null of no average treatment effect $H_{0}^{\text{weak}}: \frac{1}{N} \sum_{i=1}^{I}\sum_{j=1}^{n_{i}}(Y_{ij}(1)-Y_{ij}(0))=0$ (\citealp{neyman1923application}), which can be tested using the commonly used randomization-based difference-in-means estimator (i.e., the Neyman estimator): $T_{\text{Neyman}}(\mathbf{Z}, \mathbf{Y})=\sum_{i=1}^{I}\frac{n_{i}}{N} \hat{\tau}_{i}$, where $\hat{\tau}_{i}=\frac{1}{m_{i}}\sum_{j=1}^{n_{i}}Z_{ij}Y_{ij}-\frac{1}{n_{i}- m_{i}}\sum_{j=1}^{n_{i}}(1-Z_{ij})Y_{ij}$ (\citealp{neyman1923application}); see Appendix A.1 for details. One characteristic of our sensitivity analysis framework is that it works for both Fisher's sharp null and Neyman's weak null. In the main text, we will focus on Fisher's sharp null to explain our sensitivity analysis idea. Parallel discussions under Neyman's weak null case are presented in the online supplementary materials.

\subsection{Some basic concepts about outcome misclassification}

As discussed in Section~\ref{subsec: introduction}, in practice the measured outcomes $\mathbf{Y}^{*}=(Y_{11}^{*}, \dots, Y_{In_{I}}^{*})$ may be subject to misclassification (i.e., $\mathbf{Y}^{*} \neq \mathbf{Y})$, and a hypothesis testing result based on $\mathbf{Y}^{*}$ may differ from that based on the true outcomes $\mathbf{Y}$. In outcome misclassification and binary classification literature, the measured outcome $Y^{*}_{ij}$ is said to be a \textit{true positive} if $(Y^{*}_{ij}, Y_{ij})=(1, 1)$, a \textit{false positive} if $(Y^{*}_{ij}, Y_{ij})=(1, 0)$, a \textit{true negative} if $(Y^{*}_{ij}, Y_{ij})=(0, 0)$, or a \textit{false negative} if $(Y^{*}_{ij}, Y_{ij})=(0, 1)$. Let $TP, FP, TN$, and $FN$ denote the total number of subjects that lie in the above four categories, respectively. One of the most fundamental and widely used measures of the precision of $\mathbf{Y}^{*}$ is \textit{accuracy}, which is defined as the proportion of correct classification among $\mathbf{Y}^{*}$ under the true outcomes $\mathbf{Y}$ (denoted as $\mathcal{A}(\mathbf{Y}^{*}\mid \mathbf{Y})$):
\begin{equation*}
    \mathcal{A}(\mathbf{Y}^{*}\mid \mathbf{Y})=\frac{TP+TN}{TP+FP+TN+FN}=\frac{\sum_{i=1}^{I}\sum_{j=1}^{n_{i}}\mathbf{1}(Y_{ij}^{*}=Y_{ij})}{N}.
\end{equation*}
Note that accuracy $\mathcal{A}(\mathbf{Y}^{*}\mid \mathbf{Y})$ is a model-free and finite-population-exact concept and is therefore especially compatible with the randomization testing framework, which directly works with the finite-population study subjects in hand and does not require assuming any super-population distribution on the study subjects. 

\vspace{-0.5 cm}

\section{A model-free sensitivity analysis approach for outcome misclassification in randomization tests}\label{sec: introducing warning accuracy}
\vspace{-0.2 cm}

\subsection{Concepts and methodology}

Since outcome misclassification often exists in randomized experiments and can severely bias the downstream randomization tests, after obtaining a causal conclusion (e.g., rejecting some causal null hypothesis or not) reported by a randomization test based on the measured outcomes, a natural and fundamental question is: how many misclassified outcomes are needed at least to overturn that causal conclusion? We call this quantity the ``minimal alteration number." Or equivalently, what is the threshold such that the causal conclusion based on the measured outcomes may differ from that based on the true outcomes if the accuracy of the measured outcomes did not surpass that threshold? We call this threshold the ``warning accuracy," for which a rigorous definition is given below.
\begin{Definition}[Warning Accuracy]\label{def: warning accuracy}
Let $\mathcal{D}_{\alpha}(\mathbf{Z}, \mathbf{Y}^{*})$ and $\mathcal{D}_{\alpha}(\mathbf{Z}, \mathbf{Y})$ denote the level-$\alpha$ hypothesis testing result (rejecting the null hypothesis or not) based on the measured outcomes $\mathbf{Y}^{*}$ and that based on the true outcomes $\mathbf{Y}$ respectively, and $\mathcal{A}(\mathbf{Y}^{*}\mid \mathbf{Y})$ the accuracy of $\mathbf{Y}^{*}$ under $\mathbf{Y}$. The warning accuracy given level $\alpha$, the treatment indicators $\mathbf{Z}$, and the measured outcomes $\mathbf{Y}^{*}$, is defined as 
\begin{equation}\label{eqn: warning accuracy}
    \mathcal{WA}=\max_{\mathbf{Y}: \mathcal{D}_{\alpha}(\mathbf{Z}, \mathbf{Y}^{*})\neq \mathcal{D}_{\alpha}(\mathbf{Z}, \mathbf{Y}) }\mathcal{A}(\mathbf{Y}^{*}\mid \mathbf{Y}).
\end{equation}
Then, the corresponding minimal alteration number is $(1-\mathcal{WA})\times N$.
\end{Definition}
We here give three remarks on the warning accuracy $\mathcal{WA}$ defined in Definition~\ref{def: warning accuracy}. First, it immediately follows from Definition~\ref{def: warning accuracy} that if the actual outcome accuracy $\mathcal{A}(\mathbf{Y}^{*}\mid \mathbf{Y}) > \mathcal{WA}$, we have $\mathcal{D}_{\alpha}(\mathbf{Z}, \mathbf{Y}^{*})=\mathcal{D}_{\alpha}(\mathbf{Z}, \mathbf{Y})$ while if $\mathcal{A}(\mathbf{Y}^{*}\mid \mathbf{Y}) \leq  \mathcal{WA}$, it may happen that $\mathcal{D}_{\alpha}(\mathbf{Z}, \mathbf{Y}^{*})\neq \mathcal{D}_{\alpha}(\mathbf{Z}, \mathbf{Y})$. Therefore, other things being equal, a smaller warning accuracy $\mathcal{WA}$ indicates less sensitivity to outcome misclassification. Second, the warning accuracy is model-free and finite-population-exact as it does not require assuming the study subjects are realizations from some super-population distribution or model, which keeps the essence of randomization tests and is a free lunch added to the conventional analysis of randomization tests. Third, the concept of warning accuracy is partially inspired by some widely used concepts in the model-free sensitivity analysis literature for unmeasured confounding in observational studies, such as E-value \citep{vanderweele2017sensitivity} and sensitivity value \citep{zhao2018sensitivity}, both of which are defined as some measure of the minimal strength of unmeasured confounding needed to alter an association/causal conclusion (e.g., rejecting the null association/effect or not) of a naive analysis assuming no unmeasured confounding. Parallelly, the proposed warning accuracy is the minimal degree of outcome misclassification needed to alter a causal conclusion (e.g., rejecting the null effect or not) reported by a naive randomization test assuming no outcome misclassification. 

We now use two simple examples to illustrate why reporting the warning accuracy, as a complement of the $p$-value reported by a naive randomization test assuming no outcome misclassification, can help researchers analyze a randomized experiment (or a design-based observational study using randomization tests) subject to outcome misclassification in a more comprehensive and rigorous way.

\begin{Example}\label{exp: 1}
Consider a completely randomized experiment with one treated subject and 1000 controls. Without loss of generality, suppose that the treatment indicators $\mathbf{Z}=(1,0,\dots,0)$. Assume that the corresponding measured outcomes $\mathbf{Y}^{*}=(1,0,\dots,0)$. Then the $p$-value under Fisher's sharp null reported by Fisher's exact test based on $\mathbf{Y}^{*}$ is 1/1001<0.001, which would be considered as very strong evidence of treatment effect for many scientific journals. However, if the true outcomes $\mathbf{Y}=(0, 0, \dots, 0)$, the true $p$-value will be one (no evidence), even if $\mathbf{Y}^{*}$ and $\mathbf{Y}$ only differ by one case of misclassification. In this case, the warning accuracy is $1000/1001$, implying high sensitivity to outcome misclassification, even if the $p$-value based on $\mathbf{Y}^{*}$ is very statistically significant. 
\end{Example}
\begin{Example}\label{exp: 2}
In a more practical setting, we consider the following two stratified randomized experiments with equal sample size (=17) and the same prespecified level of 0.05 (one-sided). It is easy to check that Study 1 has a smaller one-sided $p$-value (reported by the Mantel-Haenszel test under Fisher's sharp null) but larger warning accuracy, while Study 2 has a larger one-sided $p$-value but smaller warning accuracy.

\begin{tabular}{ c c c c c c } 
   Study 1 & Stratum 1 & Stratum 2 & Stratum 3   & $p$-value & Warning Accuracy \\ 
 $\mathbf{Z}$ & (1 0 1) & (0 0 1 0 0 0) & (0 0 0 1 0 0 0 0)  & \multirow{2}{*}{1/144} & \multirow{2}{*}{16/17} \\ 
  $\mathbf{Y}^{*}$ & (1 0 1) & (0 0 1 0 0 0) & (0 0 0 1 0 0 0 0) &   &  \\
   Study 2 & Stratum 1 & Stratum 2 & Stratum 3 &  $p$-value & Warning Accuracy  \\ 
   $\mathbf{Z}$ & (1 0 0 0 0 0 0) & (1 1 0 1 0) & (0 0 1 1 0)  & \multirow{2}{*}{1/100} & \multirow{2}{*}{15/17} \\ 
  $\mathbf{Y}^{*}$ & (0 0 0 0 0 0 0) & (1 1 0 1 0) & (0 0 1 1 0)  &   & 
 \end{tabular}
\end{Example}
In summary, Example~\ref{exp: 1} shows that even a causal conclusion with a very small $p$-value reported by a randomization test can be very sensitive to outcome misclassification (i.e., high warning accuracy). Example~\ref{exp: 2} implies that a conclusion with a smaller $p$-value reported by a randomization test is not necessarily less sensitive to outcome misclassification than that with a larger $p$-value. Therefore, when outcome misclassification is a concern, reporting the warning accuracy can provide useful information about sensitivity to outcome misclassification. Such information cannot be covered by $p$-value reported by a randomization test based on measured outcomes and does not require any additional assumptions.  

Reporting warning accuracy alone for a randomization test has two limitations. First, warning accuracy is a worst-case scenario sensitivity analysis for outcome misclassification in randomization tests. While an advantage of the worst-case scenario is its universal validity, additional information or expert knowledge may be able to shed light on whether the warning accuracy is conservative in some cases. Second, warning accuracy itself cannot suggest whether a causal conclusion reported by a randomization test is especially sensitive to one of the following four types of outcome misclassification: false positives among the treated group, false negatives among the treated group, false positives among the control group, and false negatives among the control group. We propose another concept called ``sensitivity weights" to overcome these two limitations, which is built on the definition of warning accuracy and the following concept called a ``sensitive set." A sensitive set refers to a minimal (in terms of cardinality) set of subjects such that if the outcomes of those subjects were misclassified, the causal conclusion based on the measured outcomes will be overturned. Therefore, it is a collection of subjects whose outcomes being misclassified would be particularly influential to the causal conclusion. A formal definition of a sensitive set is given below.
\begin{Definition}[Sensitive Set]\label{def: sensitive sets}
Under the setting of Definition~\ref{def: warning accuracy}, let 
\begin{equation*}
   \widetilde{\mathbf{Y}}=(\widetilde{Y}_{11}, \dots, \widetilde{Y}_{In_{I}})\in \argmax_{\mathbf{Y}: \mathcal{D}_{\alpha}(\mathbf{Z}, \mathbf{Y}^{*})\neq \mathcal{D}_{\alpha}(\mathbf{Z}, \mathbf{Y}) }\mathcal{A}(\mathbf{Y}^{*}\mid \mathbf{Y})
\end{equation*}
be an optimal solution to the optimization problem (\ref{eqn: warning accuracy}) associated with the definition of warning accuracy. Then $\mathcal{S}=\{ij: \widetilde{Y}_{ij}\neq Y_{ij}^{*}\}$ is called a sensitive set.
\end{Definition}
In practice, there may exist more than one sensitive set as the solution to the optimization problem involved in Definition~\ref{def: warning accuracy} may not be unique. However, as we will show in Section~\ref{sec: computing}, these different sensitive sets can be transformed into each other by some composition of within-strata and between-strata permutations and share the same quantities called \textit{sensitivity wights}--the key to answer \textbf{Q2} and \textbf{Q3} described in Section~\ref{subsec: contributions}.

\begin{Definition}[Sensitivity Weights]\label{def: sensitivity weights}
Under the setting of Definition~\ref{def: warning accuracy}, let $\mathcal{S}$ be a sensitive set. Then there is a set of sensitivity weights defined as the following $2 \times 2$ table:
\begin{center}
    \begin{tabular}{ccc} 
  \hline
  Sensitivity Weights & False Positives & False Negatives   \\
  \hline
   Treated   & $W_{T}^{FP}$ & $W_{T}^{FN}$\\
   Control & $W_{C}^{FP}$ & $W_{C}^{FN}$   \\
 \hline
 \end{tabular}
\end{center}
where $W_{T}^{FP}$, $W_{T}^{FN}$, $W_{C}^{FP}$, and $W_{C}^{FN}$ denote the sensitivity weight of false positives in the treated group, that of false negatives in the treated group, that of false positives in the control group, and that of false negatives in the control group respectively, defined as 
\begin{align*}
    &W_{T}^{FP}=\frac{|\{ij: Z_{ij}=1, Y_{ij}^{*}=1, \widetilde{Y}_{ij}=0\}|}{|\mathcal{S}|}, \quad W_{T}^{FN}=\frac{|\{ij: Z_{ij}=1, Y_{ij}^{*}=0, \widetilde{Y}_{ij}=1\}|}{|\mathcal{S}|},\\
    &W_{C}^{FP}=\frac{|\{ij: Z_{ij}=0, Y_{ij}^{*}=1, \widetilde{Y}_{ij}=0\}|}{|\mathcal{S}|}, \quad W_{C}^{FN}=\frac{|\{ij: Z_{ij}=0, Y_{ij}^{*}=0, \widetilde{Y}_{ij}=1\}|}{|\mathcal{S}|}.
\end{align*}
Note that we have $W_{T}^{FP}+W_{T}^{FN}+W_{C}^{FP}+W_{C}^{FN}=1$.
\end{Definition}
The rationale of sensitivity weights is straightforward: since the causal conclusion is particularly sensitive to potential cases of outcome misclassification among the subjects in a sensitive set, then the proportion of each of the four types of outcome misclassification (i.e., false positives/negatives in the treated/control group) within a sensitive set offers a sensible quantification about if the causal conclusion is in particular sensitive to a certain type of outcome misclassification. For example, in Example~\ref{exp: 1}, it is clear that the sensitivity weight of a false positive in the treated group is the dominant term according to Definition~\ref{def: sensitivity weights}, suggesting that the causal conclusion is especially sensitive to a false positive in the treated group and therefore should be given priority when conducting a validation study. 

\subsection{Practical guidance}

Putting all the concepts developed above together, to address \textbf{Q1} described in Section~\ref{subsec: contributions}, after reporting the $p$-value using a naive randomization test assuming no outcome misclassification, researchers can then report the warning accuracy under the prespecified significance level. A low warning accuracy suggests a causal conclusion's high robustness to outcome misclassification. If the warning accuracy is relatively high instead, researchers should report the sensitivity weights and check if the dominant term among the four types of outcome misclassification (false positives/negatives in the treated/control group) agrees with that based on prior information and/or expert knowledge or not (we will illustrate such procedure in detail in Section~\ref{sec: PCPT analysis}). If this is the case, the causal conclusion drawn from the measured outcomes may be misleading even if the unknown actual outcome accuracy equals or is close to that high warning accuracy. Otherwise, we would expect the actual accuracy needed to overturn the causal conclusion based on the measured outcomes to be lower, or even much lower, than that high warning accuracy. To address \textbf{Q2} described in Section~\ref{subsec: contributions}, we report the four sensitivity weights and observe if there is any dominant term among the four sensitivity weights. Such a term would be a strong sign that the causal conclusion reported by a randomization test based on the measured outcomes is particularly sensitive to that type of outcome misclassification. To address \textbf{Q3} described in Section~\ref{subsec: contributions}, if the warning accuracy is relatively high and the dominant term among the four sensitivity weights is expected in practice (i.e., the study is sensitive to outcome misclassification), then researchers may want to select a subset of study subjects for outcome validation. When choosing a validation subset, it makes sense to give priority to (i) the type of outcome misclassification with the dominant sensitivity weight and (ii) the subjects that belong to a sensitive set. 

\vspace{-0.5 cm}

\section{Computing warning accuracy and related quantities}\label{sec: computing}

\vspace{-0.2 cm}

\subsection{The original problem formulation and the ``curse of symmetry"}\label{subsec: original problem}

When the sample size $N$ is large, it is typically infeasible to calculate warning accuracy by hand as in Section~\ref{sec: introducing warning accuracy}. A general strategy for tackling a computationally extensive problem involving integers, such as calculating warning accuracy (\ref{eqn: warning accuracy}), is to appropriately formulate the problem into an integer program and apply a state-of-the-art optimization solver. To illustrate the general idea, we first consider testing $H_{0}^{\text{sharp}}$ with binary outcomes using the routinely used Mantel-Haenszel test, which reduces to Fisher's exact test when $I=1$ and McNemar's test when $n_{i}=2$ for all $i$. If $H_{0}^{\text{sharp}}$ was rejected by the Mantel Haenszel test based on $\mathbf{Y}^{*}$, i.e., if $\frac{[T_{\text{M-H}}(\mathbf{Z}, \mathbf{Y}^{*})-E\{T_{\text{M-H}}(\mathbf{Z}, \mathbf{Y}^{*})\} ]^{2}}{\text{Var}\{T_{\text{M-H}}(\mathbf{Z}, \mathbf{Y}^{*})\} }>\chi^{2}_{1, 1-\alpha}$, by Definition~\ref{def: warning accuracy}, the warning accuracy is the optimal value of the following integer quadratically constrained linear program (IQCLP):
\begin{equation*}
     \begin{split}
        \underset{\mathbf{Y}\in \{0,1\}^{N} } {\text{maximize}} \quad &\frac{1}{N}\sum_{i=1}^{I}\sum_{j=1}^{n_{i}}Y_{ij}^{*}Y_{ij}+\frac{1}{N}\sum_{i=1}^{I}\sum_{j=1}^{n_{i}}(1-Y_{ij}^{*})(1-Y_{ij}) \quad \quad (\text{P}0)\\
         \text{subject to}\quad & [T_{\text{M-H}}(\mathbf{Z}, \mathbf{Y})-E\{T_{\text{M-H}}(\mathbf{Z}, \mathbf{Y})\}]^{2}-\chi^{2}_{1, 1-\alpha} \cdot \text{Var}\{T_{\text{M-H}}(\mathbf{Z}, \mathbf{Y})\}\leq 0.
     \end{split}
 \end{equation*}
 The objective function of (P0) comes from Definition~\ref{def: warning accuracy} and a simple observation that $\sum_{i=1}^{I}\sum_{j=1}^{n_{i}}\mathbf{1}(Y_{ij}^{*}=Y_{ij})=\sum_{i=1}^{I}\sum_{j=1}^{n_{i}}Y_{ij}^{*}Y_{ij}+\sum_{i=1}^{I}\sum_{j=1}^{n_{i}}(1-Y_{ij}^{*})(1-Y_{ij})$, which is a linear function in $\mathbf{Y}$ given $\mathbf{Y}^{*}$. The inequality constraint of (P0) comes from the fact that $\mathcal{D}_{\alpha}(\mathbf{Z}, \mathbf{Y})\neq \mathcal{D}_{\alpha}(\mathbf{Z}, \mathbf{Y}^{*})=``Reject"$ if and only if $\frac{[T_{\text{M-H}}(\mathbf{Z}, \mathbf{Y})-E\{T_{\text{M-H}}(\mathbf{Z}, \mathbf{Y})\} ]^{2}}{\text{Var}\{T_{\text{M-H}}(\mathbf{Z}, \mathbf{Y})\} }\leq \chi^{2}_{1, 1-\alpha}$, which can be rewritten as a quadratic constraint in $\mathbf{Y}$ as in (P0). Instead, if $H_{0}^{\text{sharp}}$ fails to be rejected by the Mantel-Haenszel test based on $\mathbf{Y}^{*}$, to compute the warning accuracy, we just need to solve a simple variant of (P0) with replacing the ``$\leq 0$" with the ``$\geq 0$" in the constraint. Typically, conducting a sensitivity analysis (e.g., calculating warning accuracy) or a validation study (whose study design can be guided by learning sensitivity weights) makes more sense if the study's null hypothesis was rejected (i.e., some treatment effect was detected) by a primary analysis (\citealp{rosenbaum2002observational}) based on measured outcomes, so in the rest of this paper we focus on the ``$\leq 0$" constraint case. For calculating the warning accuracy under Neyman's weak null, we just need to replace the inequality constraint in (P0) with $\{T_{\text{Neyman}}(\mathbf{Z}, \mathbf{Y})\}^{2}-\chi^{2}_{1,1-\alpha} \cdot \widehat{\text{Var}}\{T_{\text{Neyman}}(\mathbf{Z}, \mathbf{Y})\} \leq 0$ (or $\geq 0$), in which $\widehat{\text{Var}}\{T_{\text{Neyman}}(\mathbf{Z}, \mathbf{Y})\}$ is a conservative variance estimator for $T_{\text{Neyman}}$ (see Appendix A.1 for details). This is still a quadratic constraint in $\mathbf{Y}$. See Web Appendix A.3 for details.

Before discussing the computational issues in the next paragraph, we here give two important remarks: 1) In this article, to illustrate our general sensitivity analysis framework, we implemented the commonly used Mantel-Haenszel test (for sharp null) and the Neyman estimator (for weak null). To implement other randomization tests, say, some chosen randomization test statistics $T_{*}(\mathbf{Z}, \mathbf{Y})$, we just need to replace the inequality constraint in (P0) with $\{T_{*}(\mathbf{Z}, \mathbf{Y})\}^{2}-\chi^{2}_{1,1-\alpha} \cdot \widehat{\text{Var}}\{T_{*}(\mathbf{Z}, \mathbf{Y})\} \leq 0$ (or $\geq 0$), as long as the finite-population central limit theorem holds for $T_{*}(\mathbf{Z}, \mathbf{Y})$. 2) Our framework is flexible enough to incorporate additional prior information about outcome misclassification by imposing additional constraints into (P0). For example, a bound on the differences between the four types of outcome misclassification (i.e., false positives/negatives among the treated/control group) can be implemented by adding additional linear constraints into (P0).
 
The original problem formulation (P0) seems straightforward and natural. However, this seemingly reasonable integer program formulation can easily make the computation infeasible because of the so-called ``curse of symmetry" (\citealp{margot2010symmetry}), as will be explained further below. For clarity of the notation, without loss of generality, in Section~\ref{subsec: original problem}, we temporarily re-organize subjects such that for each stratum $i$, we have: (i) $Z_{ij_{1}}\leq Z_{ij_{2}}$ as long as $j_{1}\leq j_{2}$; (ii) $Y_{ij_{1}}^{*}\leq Y_{ij_{2}}^{*}$ as long as $j_{1}\leq j_{2}$ and $Z_{ij_{1}}=Z_{ij_{2}}$. Following the typical notation in group theory (\citealp{scott2012group}), we let $(ij, ij^{\prime})$ denote the permutation of the index set $\mathcal{I}=\{ 11, 12, \dots, In_{I}\}$ such that index $ij$ exchanges with index $ij^{\prime}$ while all other indexes remain the same. Define the following permutation group $G_{\text{within}}$ over $\mathcal{I}: G_{\text{within}}=\{$All possible compositions of any $(ij, ij^{\prime})$ s.t. $Y_{ij}^{*}=Y_{ij^{\prime}}^{*}$ and $Z_{ij}=Z_{ij^{\prime}}\}$. Let $g\mathbf{Y}$ denote that the permutation $g$ acts on the indexes in $\mathbf{Y}: g\mathbf{Y}=g(Y_{11}, \dots, Y_{In_{I}})=(Y_{g(11)}, \dots, Y_{g(In_{I})})$. It is clear that for any $g \in G_{\text{within}}$, we have $\mathcal{A}(\mathbf{Y}^{*}\mid g\mathbf{Y})=\mathcal{A}(\mathbf{Y}^{*}\mid \mathbf{Y})$ and $\mathcal{D}_{\alpha}(\mathbf{Z}, g\mathbf{Y})= \mathcal{D}_{\alpha}(\mathbf{Z}, \mathbf{Y})$, i.e., the integer program (P0) is invariant under any permutation $g \in G_{\text{within}}$. We call this property \textit{within-strata symmetry}. Then for each stratum $i$, let $\mathbf{\Lambda}_{i}=(\Lambda_{i}^{00}, \Lambda_{i}^{01}, \Lambda_{i}^{10}, \Lambda_{i}^{11})=(\sum_{j=1}^{n_{i}}(1-Z_{ij})(1-Y_{ij}^{*}), \sum_{j=1}^{n_{i}}(1-Z_{ij})Y_{ij}^{*}, \sum_{j=1}^{n_{i}}Z_{ij}(1-Y_{ij}^{*}), \sum_{j=1}^{n_{i}}Z_{ij}Y_{ij}^{*})$ denote the measured $2\times 2$ table of stratum $i$. Let $S$ denote the number of unique $2\times 2$ tables among $\{\mathbf{\Lambda}_{i}, i=1,\dots, I\}$, and let $P_{s}$ denote the number of strata with the measured $2\times 2$ tables equalling the $s$-th unique table $\mathbf{\Lambda}_{[s]}=(\Lambda_{[s]}^{00}, \Lambda_{[s]}^{01}, \Lambda_{[s]}^{10}, \Lambda_{[s]}^{11}), s=1,\dots, S$, therefore $\sum_{s=1}^{S}P_{s}=I$. For two strata $i$ and $i^{\prime}$ with $n_{i}=n_{i^{\prime}}$, we let $(i, i^{\prime})=(i1, i^{\prime}1)\dots (in_{i}, i^{\prime}n_{i^{\prime}})$ denote the permutation of $\mathcal{I}$ such that stratum $i$'s indexes element-wisely exchange with stratum $i^{\prime}$'s indexes while all other indexes remain the same. Define the following permutation group $G_{\text{between}}=\{\text{All possible compositions of any } (i, i^{\prime}) \text{ s.t. } \mathbf{\Lambda}_{i}=\mathbf{\Lambda}_{i^{\prime}} \}$ over $\mathcal{I}$. It is clear that for any $g \in G_{\text{between}}$, we also have $\mathcal{A}(\mathbf{Y}^{*}\mid g\mathbf{Y})=\mathcal{A}(\mathbf{Y}^{*}\mid \mathbf{Y})$ and $\mathcal{D}_{\alpha}(\mathbf{Z}, g\mathbf{Y})= \mathcal{D}_{\alpha}(\mathbf{Z}, \mathbf{Y})$, i.e., the integer program (P0) is invariant under any permutation $g \in G_{\text{between}}$. We call such property as \textit{between-strata symmetry}. An illustration of the two types of symmetry (i.e., within-strata and between-strata symmetry) is given in Table~\ref{tab: two types of symmetry}. 
 
 \begin{table}[b]
\caption{Illustration of the two types of symmetry: between-strata symmetry (e.g., stratum 1 and stratum 2) and within-strata symmetry (e.g., subject 1 and subject 2 in stratum 3).}
     \label{tab: two types of symmetry}
\begin{minipage}{0.4\linewidth}
     \centering
    \begin{tabular}{c| c c } 
   & $Y^{*}=1$ & $Y^{*}=0$ \\ 
 \hline
 \multirow{2}{*}{$Z=1$} & \semaphore{black}{black} & \semaphore{black}{shadecolor} \\ 
 \multirow{2}{*}{$Z=0$} & \semaphore{shadecolor}{black} & \semaphore{shadecolor}{shadecolor} \\
 \end{tabular}

\bigskip

(a) Four types of study subjects.

\end{minipage}\hfill
\begin{minipage}{0.6\linewidth}

     \centering
     \begin{tabular}{c| c c c} 
   & Subject 1 & Subject 2 & Subject 3 \\ 
 \hline
 \multirow{2}{*}{Stratum 1} & \semaphore{black}{black} & \semaphore{black}{shadecolor} & \semaphore{shadecolor}{shadecolor} \\ 
 \multirow{2}{*}{Stratum 2} & \semaphore{black}{black} & \semaphore{black}{shadecolor} & \semaphore{shadecolor}{shadecolor} \\
  \multirow{2}{*}{Stratum 3} & \semaphore{black}{shadecolor} & \semaphore{black}{shadecolor}  & \semaphore{shadecolor}{shadecolor} 
 \end{tabular}
 
 \bigskip

(b) Two types of symmetry.

\end{minipage}
\end{table}
 
 Putting the above discussions together, define the permutation group $G=\{$All possible compositions of elements from $G_{\text{within}}$ and $G_{\text{between}}\}$. For any $g \in G$, the integer program (P0) is invariant under permutation $g$. Note that \begin{equation*}
      |G|=|G_{\text{between}}|\times |G_{\text{within}}|=\prod_{s=1}^{S}P_{s}!\times \prod_{i=1}^{I}\Lambda^{00}_{i}!\Lambda^{01}_{i}!\Lambda^{10}_{i}!\Lambda^{11}_{i}!,
 \end{equation*}
which is an extremely large number if $N$ is large, indicating an extremely high degree of symmetry of the integer program (P0) and resulting in computational infeasibility of solving (P0) due to the so-called ``curse of symmetry," which refers to a general fact that an integer program is typically computationally infeasible if its variables can be permuted in many ways (e.g., $|G|$ is large) without changing the problem structure (\citealp{margot2010symmetry}). 

In addition to the computational challenge, another implication of the above arguments is that, as mentioned in Section~\ref{sec: introducing warning accuracy}, a sensitive set (i.e., an optimal solution to the integer program (P0)) itself is not an intrinsic concept because for a given sensitive set, its transformation under some between-strata permutations (as in $G_{\text{between}}$) and/or within-strata permutations (as in $G_{\text{within}}$) is still a sensitive set. However, a simple but important observation is that these different sensitive sets have the same sensitivity weights. As mentioned in Section~\ref{sec: introducing warning accuracy}, this motivates us to introduce an intrinsic concept associated with a sensitive set -- sensitivity weights, which is invariant under any permutations in $G_{\text{between}}$ and  $G_{\text{within}}$ and therefore provides an intrinsic quantification of a causal conclusion's relative sensitivity to the four different types of outcome misclassification.

\subsection{Two types of randomization designs and an adaptive reformulation strategy}\label{subsec: adaptive}

In Section~\ref{subsec: adaptive}, we propose a general strategy to solve the ``curse of symmetry" encountered when calculating the warning accuracy according to the original problem formulation (P0). The core idea of our strategy is to reformulate the integer program (P0) with respect to an intrinsic characteristic of various randomization designs--whether within-strata symmetry dominates between-strata symmetry for that randomization design or vice versa. Specifically, we classify many common randomization designs into the following two types: 1) \textbf{Type I randomization designs}: those in which within-strata symmetry dominates between-strata symmetry (i.e., $|G_{\text{within}}|\gg|G_{\text{between}}|$). This class of randomization designs includes some commonly used randomization designs (including randomized experiments and observational studies adopting randomization-based inference), such as completely randomized experiments (for which $|G_{\text{between}}|=1$) and stratified/blocked randomized experiments or observational studies with most strata/blocks being large. 2) \textbf{Type II randomization designs}: those in which between-strata symmetry dominates within-strata symmetry (i.e., $|G_{\text{between}}|\gg|G_{\text{within}}|$). This class of randomization designs includes some widely used randomization designs such as paired randomized experiments or pair-matched observational studies (for which $|G_{\text{within}}|=1$), randomized experiments or observational studies with multiple controls, randomized experiments or observational studies with variable controls, finely stratified experiments or observational studies with full matching, and stratified/blocked randomized experiments or observational studies with most strata/blocks being small.

\subsubsection{Type I: within-strata symmetry dominates between-strata symmetry}\label{subsubsec: type I}

We first show how to reformulate the integer program (P0) for type I randomization designs. By independence of treatment assignments between strata, the definition of $\mathcal{A}(\mathbf{Y}\mid \mathbf{Y}^{*})$, and the definition of the Mantel-Haenszel test, the accuracy (the objective function of (P0)) can be determined by the $I$ $2\times 2$ tables $\{\sum_{j=1}^{n_{i}}\mathbf{1}(Y_{ij}^{*}=b, Y_{ij}=c): b, c \in \{0,1\}\}$ ($i=1,\dots,I$) and the constraint's feasibility (i.e., the Mantel-Haenszel test fails to reject $H_{0}^{\text{sharp}}$ based on $\mathbf{Y}$) can be determined by the $I$ $2\times 2$ tables $\{\sum_{j=1}^{n_{i}}\mathbf{1}(Z_{ij}=a, Y_{ij}=c): a, c \in \{0,1\}\}$ ($i=1,\dots,I$). Therefore, the integer program (P0) can be determined by the $I$ $2\times 2 \times 2$ tables $\{\sum_{j=1}^{n_{i}}\mathbf{1}(Z_{ij}=a, Y_{ij}^{*}=b, Y_{ij}=c): a, b, c \in \{0,1\}\}$ ($i=1,\dots,I$). Because $\mathbf{Z}$ and $\mathbf{Y}^{*}$ have been observed, for each stratum $i$, the $i$-th $2\times 2 \times 2$ table only has the following four degrees of freedom: $\Upsilon_{i}^{00}=\sum_{j=1}^{n_{i}}\mathbf{1}(Z_{ij}=0, Y_{ij}^{*}=0, Y_{ij}=1)$, $\Upsilon_{i}^{01}=\sum_{j=1}^{n_{i}}\mathbf{1}(Z_{ij}=0, Y_{ij}^{*}=1, Y_{ij}=1)$, $\Upsilon_{i}^{10}=\sum_{j=1}^{n_{i}}\mathbf{1}(Z_{ij}=1, Y_{ij}^{*}=0, Y_{ij}=1)$, and $\Upsilon_{i}^{11}=\sum_{j=1}^{n_{i}}\mathbf{1}(Z_{ij}=1, Y_{ij}^{*}=1. Y_{ij}=1)$. Let $\mathbf{\Upsilon}=(\Upsilon_{1}^{00}, \Upsilon_{1}^{01}, \Upsilon_{1}^{10}, \Upsilon_{1}^{11}, \dots, \Upsilon_{I}^{00}, \Upsilon_{I}^{01}, \Upsilon_{I}^{10}, \Upsilon_{I}^{11})\in  \mathbb{Z}^{4I}$ and $\breve{\Upsilon}_{i}=\Upsilon_{i}^{00}+\Upsilon_{i}^{01}+\Upsilon_{i}^{10}+\Upsilon_{i}^{11}$, then (P0) can be reformulated as the following IQCLP (P1):
\begin{equation*}
     \begin{split}
        \underset{\mathbf{\Upsilon}\in  \mathbb{Z}^{4I}} {\text{max}} \quad &\frac{1}{N}\sum_{i=1}^{I}(\Upsilon_{i}^{01}+\Upsilon_{i}^{11}-\Upsilon_{i}^{00}-\Upsilon_{i}^{10})+\frac{1}{N}\sum_{i=1}^{I}\sum_{j=1}^{n_{i}}(1-Y_{ij}^{*}) \quad \quad (\text{P}1)\\
         \text{s.t.}\quad & \Big\{ \sum_{i=1}^{I}(\Upsilon_{i}^{10}+\Upsilon_{i}^{11})-\sum_{i=1}^{I}\frac{m_{i}}{n_{i}} \breve{\Upsilon}_{i} \Big\}^{2}- \chi^{2}_{1, 1-\alpha} \cdot \sum_{i=1}^{I}\frac{m_{i} \breve{\Upsilon}_{i}  (n_{i}-\breve{\Upsilon}_{i})(n_{i}-m_{i})}{n_{i}^{2}(n_{i}-1)}\leq 0, \\
        &0 \leq \Upsilon_{i}^{00} \leq \sum_{j=1}^{n_{i}}(1-Z_{ij})(1-Y_{ij}^{*}), \quad \ \ 0 \leq \Upsilon_{i}^{01} \leq \sum_{j=1}^{n_{i}}(1-Z_{ij})Y_{ij}^{*}, \\
        &0 \leq \Upsilon_{i}^{10} \leq \sum_{j=1}^{n_{i}}Z_{ij}(1-Y_{ij}^{*}), \qquad \qquad  0 \leq \Upsilon_{i}^{11} \leq \sum_{j=1}^{n_{i}}Z_{ij}Y_{ij}^{*}, \qquad \forall i. \\ 
     \end{split}
 \end{equation*}
 
Note that in (P1), in addition to rewriting the objective function and constraint of (P0) in terms of new decision variables $\mathbf{\Upsilon}$, we also add the bounding constraints for $\mathbf{\Upsilon}$. It is clear that there is no within-strata symmetry anymore in (P1), which was previously a major source of symmetry in the original formulation (P0) for type I randomization designs. The above argument also works for the case of testing Neyman's weak null $H_{0}^{\text{weak}}$ with type I randomization designs; see Web Appendix B.3 for details. In Web Appendix C, we showed how to calculate sensitivity weights and a collection of sensitive sets after solving (P1).

\subsubsection{Type II: between-strata symmetry dominates within-strata symmetry}\label{subsubsec: type II}

Our strategy of reformulating the integer program (P0) for type II randomization designs is inspired by both the observations discussed in Section~\ref{subsubsec: type I} and an idea from \citet{fogarty2016discrete}. Specifically, in \citet{fogarty2016discrete}, to find the worst-case variance estimator for the average treatment effect with binary outcomes after matching, instead of manipulating the potential outcomes based on the treatment indicators and observed outcomes, the authors proposed to list all possible $2 \times 2 \times 2$ tables (in terms of treatment indicators, observed outcomes, and potential outcomes) based on observed data and manipulate the number of each possible $2 \times 2 \times 2$ table that enter into randomization inference. Although in this paper we are studying a totally different problem, we can combine this idea and the arguments in Section~\ref{subsubsec: type I} to reformulate the integer program (P0) for type II randomization designs. 

A high-level summary of the idea is: for type II randomization designs, there are many ``words" (strata) but not many ``vocabularies" (all possible $2 \times 2 \times 2$ tables in terms of binary treatment indicators, binary measured outcomes, and binary true outcomes, given the observed data). Therefore, instead of directly manipulating the potential $2\times 2 \times 2$ table for each stratum as in (P1), we first create a ``dictionary" that lists all possible unique $2 \times 2 \times 2$ tables based on measured $2 \times 2$ tables (in terms of treatment indicators and measured outcomes) and then manipulate the total number of strata that have certain $2 \times 2 \times 2$ table. We call this strategy the ``dictionary method." Specifically, for the $s$-th unique $2 \times 2$ table $\mathbf{\Lambda}_{[s]}=(\Lambda_{[s]}^{00}, \Lambda_{[s]}^{01}, \Lambda_{[s]}^{10}, \Lambda_{[s]}^{11}), s=1,\dots, S$, let $\widetilde{n}_{s}$ and $\widetilde{m}_{s}$ denote its number of total study subjects and its number of treated subjects. Moreover, for the $s$-th unique $2 \times 2$ table $\mathbf{\Lambda}_{[s]}$, there are $\widetilde{N}_{s}=(\Lambda_{[s]}^{00}+1) (\Lambda_{[s]}^{01}+1)(\Lambda_{[s]}^{10}+1)(\Lambda_{[s]}^{11}+1)$ possible unique $2\times 2 \times 2$ tables $\{\sum_{j=1}^{\widetilde{n}_{s}}\mathbf{1}(Z_{ij}=a, Y_{ij}^{*}=b, Y_{ij}=c): a, b, c \in \{0,1\}\}$ ($s=1,\dots,S$). Let $d_{sp}$ be the number of the $p$-th unique $2\times 2 \times 2$ table $\Delta_{sp}$ for the $s$-th unique table $\mathbf{\Lambda}_{[s]}$, $s=1,\dots,S$ and $p=1, \dots, \widetilde{N}_{s}$. Since $\mathbf{Z}$ and $\mathbf{Y}^{*}$ are given, each $2\times 2 \times 2$ table $\Delta_{sp}$ can be uniquely determined by four numbers $\Delta_{sp}^{00}=\sum_{j=1}^{\widetilde{n}_{s}}\mathbf{1}(Z_{ij}=0, Y_{ij}^{*}=0, Y_{ij}=1)$, $\Delta_{sp}^{01}=\sum_{j=1}^{\widetilde{n}_{s}}\mathbf{1}(Z_{ij}=0, Y_{ij}^{*}=1, Y_{ij}=1)$, $\Delta_{sp}^{10}=\sum_{j=1}^{\widetilde{n}_{s}}\mathbf{1}(Z_{ij}=1, Y_{ij}^{*}=0, Y_{ij}=1)$, and $\Delta_{sp}^{11}=\sum_{j=1}^{\widetilde{n}_{s}}\mathbf{1}(Z_{ij}=1, Y_{ij}^{*}=1, Y_{ij}=1)$. Let $\breve{\Delta}_{sp}= \Delta_{sp}^{00}+\Delta_{sp}^{01}+\Delta_{sp}^{10}+\Delta_{sp}^{11}$, then we can reformulate the integer program (P0) as the following IQCLP (P2):
\begin{equation*}
     \begin{split}
        \underset{d_{sp}\in \mathbb{Z}, d_{sp}\geq 0}{\text{max}} \quad & \frac{1}{N}\sum_{s=1}^{S}\sum_{p=1}^{\widetilde{N}_{s}}d_{sp}(\Delta_{sp}^{01}+\Delta_{sp}^{11}-\Delta_{sp}^{00}-\Delta_{sp}^{10})+\frac{1}{N}\sum_{i=1}^{I}\sum_{j=1}^{n_{i}}(1-Y_{ij}^{*}) \quad \quad (\text{P}2)\\
         \text{s.t. }\quad  & \Big\{ \sum_{s=1}^{S}\sum_{p=1}^{\widetilde{N}_{s}}d_{sp}(\Delta_{sp}^{10}+\Delta_{sp}^{11})-\sum_{s=1}^{S}\sum_{p=1}^{\widetilde{N}_{s}}d_{sp}\Big (\frac{\widetilde{m}_{s}}{\widetilde{n}_{s}}\cdot \breve{\Delta}_{sp}\Big) \Big\}^{2}\\
   &\quad \quad \quad \quad \quad \quad  - \chi^{2}_{1, 1-\alpha} \cdot \sum_{s=1}^{S}\sum_{p=1}^{\widetilde{N}_{s}}d_{sp}\frac{\widetilde{m}_{s} \breve{\Delta}_{sp} (\widetilde{n}_{s}-\breve{\Delta}_{sp}) (\widetilde{n}_{s}-\widetilde{m}_{s})}{\widetilde{n}_{s}^{2}(\widetilde{n}_{s}-1)}\leq 0, \\
        & \sum_{p=1}^{\widetilde{N}_{s}} d_{sp} = P_{s}. \quad \forall s 
     \end{split}
 \end{equation*}
Note that in (P2), $(\Delta_{sp}^{00}, \Delta_{sp}^{01}, \Delta_{sp}^{10}, \Delta_{sp}^{11})$ are fixed numbers (i.e., ``vocabularies" listed in a ``dictionary") and the decision variables are $d_{sp}$, i.e., the total number of strata that take the unique $2 \times 2 \times 2$ table $(\Delta_{sp}^{00}, \Delta_{sp}^{01}, \Delta_{sp}^{10}, \Delta_{sp}^{11})$. It is clear that there is neither within-strata symmetry nor between-strata symmetry in (P2) and, therefore, surpasses the formulation (P1) in terms of erasing symmetry. However, the dimension of the decision variables in (P2) is $\sum_{s=1}^{S}\widetilde{N}_{s}=\sum_{s=1}^{S}(\Lambda_{[s]}^{00}+1) (\Lambda_{[s]}^{01}+1)(\Lambda_{[s]}^{10}+1)(\Lambda_{[s]}^{11}+1)$, which is typically very high for type I randomization designs, but is typically not high for type II randomization. Therefore, for type II randomization designs, it is appropriate to reformulate (P0) as (P2). Whereas, for type I randomization designs, for which within-strata symmetry is the main concern, it is more appropriate to reformulate (P0) as (P1). 

The above argument also works for the case of testing Neyman's weak null with type II randomization designs; see Web Appendix B.4 for details. A method of calculating sensitivity weights and a collection of sensitive sets after solving (P2) is given in Web Appendix C.
 
\subsection{Simulation studies}\label{subsec: simulation}

We conduct simulations to study the computational efficiency of the adaptive reformulation strategy proposed in Section~\ref{subsec: adaptive}, with a modern optimization solver \textsf{Gurobi} \citep{gurobi}. We also gain some insights into how warning accuracy and sensitivity weights vary with the effect size and sample size. Note that the data-generating processes described below are only intended for automatically generating data sets for simulations, as our framework works for finite-population data sets and does not require assuming some specific data-generating model. We investigate both type I and type II randomization designs by considering the following two simulation scenarios:
\begin{itemize}
   \item \textbf{Simulation Scenario 1 (for Type I randomization designs):} We consider a stratified randomized experiment (or a stratified observational study adopting randomization tests) with $I=40$ or $200$ strata. We let $\mathcal{U}(A)$ denote the uniform distribution over the set $A$. In each independent simulation run, for each $i=1,\dots, I$ we randomly draw $m_{i}$ from $\mathcal{U}(\{10, 11, \dots, 40\})$ and then randomly draw $n_{i}-m_{i}$ from $\mathcal{U}(\{ 10, 11, \dots, 40\})$. The expected total number of study subjects $E(N)=\{E(m_{i})+E(n_{i}-m_{i})\}\cdot I=50I=2000$ or $10,000$.
    
   \item \textbf{Simulation Scenario 2 (for Type II randomization designs):} We consider a finely stratified randomized experiment (or a matched observational study with full matching) with $I=400$ or $2000$ strata. In each independent simulation run, for each $i=1,\dots, I$ we randomly generate $m_{i}$ and $n_{i}-m_{i}$ based on the following procedure: we first set $\min \{ m_{i}, n_{i}-m_{i}\}=1$ and draw $\max \{m_{i}, n_{i}-m_{i}\}$ from $\mathcal{U}(\{ 1, 2, \dots, 7\})$ and then randomly assign $\min \{ m_{i}, n_{i}-m_{i}\}$ and $\max \{m_{i}, n_{i}-m_{i}\}$ to $m_{i}$ and $n_{i}-m_{i}$. Then we have $E(N)=\{E(m_{i})+E(n_{i}-m_{i})\}\cdot I=5I=2000$ or $10,000$.
\end{itemize}

\begin{table}[b]
\scriptsize
\centering 
\caption{The average computation time (in seconds), warning accuracy $\mathcal{WA}$ and sensitivity weights $(W_{T}^{FP}, W_{T}^{FN}, W_{C}^{FP}, W_{C}^{FN})$ of different sets of $(E(N), p_{0}, p_{1})$.}
\label{tab: simulation for sharp null}
\resizebox{\textwidth}{!}{\begin{tabular}{ccccccccccccc}
\hline
\multicolumn{13}{c}{\textbf{Type I Randomization Designs (Simulation Scenario 1)}} \\
\hline
\multirow{2}{*}{$p_{0}=0.3$} & \multicolumn{6}{c}{$E(N)=2000$} & \multicolumn{6}{c}{$E(N)=10,000$}\\
\cmidrule(r){2-7} \cmidrule(r){8-13} 
& Time & $\mathcal{WA}$ & $W_{T}^{FP}$ & $W_{T}^{FN}$ & $W_{C}^{FP}$ & $W_{C}^{FN}$ & Time & $\mathcal{WA}$ & $W_{T}^{FP}$ & $W_{T}^{FN}$ & $W_{C}^{FP}$ & $W_{C}^{FN}$ \\
\hline
$p_{1}=0.4$ & 0.15 s & 0.98 & 0.33 & 0.00 & 0.00  & 0.67 & 3.67 s & 0.97 & 0.35 & 0.00 & 0.00 & 0.65 \\
$p_{1}=0.6$ & 0.17 s & 0.91 & 0.46 & 0.00 & 0.00 & 0.54 & 3.73 s & 0.90 & 0.46 & 0.00 & 0.00  & 0.54  \\
$p_{1}=0.8$ & 0.18 s & 0.83 & 0.54 & 0.00 & 0.00 & 0.46  & 3.79 s & 0.82 & 0.54 & 0.00  & 0.00 & 0.46  \\
\hline
\multirow{2}{*}{$p_{0}=0.6$} & \multicolumn{6}{c}{$E(N)=2000$} & \multicolumn{6}{c}{$E(N)=10,000$}\\
\cmidrule(r){2-7} \cmidrule(r){8-13} 
& Time & $\mathcal{WA}$ & $W_{T}^{FP}$ & $W_{T}^{FN}$ & $W_{C}^{FP}$ & $W_{C}^{FN}$ & Time & $\mathcal{WA}$ & $W_{T}^{FP}$ & $W_{T}^{FN}$ & $W_{C}^{FP}$ & $W_{C}^{FN}$ \\
\hline
$p_{1}=0.7$ & 0.15 s & 0.98 & 0.68 & 0.00 & 0.00 & 0.32 & 3.66 s & 0.97 & 0.66 & 0.00 & 0.00 & 0.34 \\
$p_{1}=0.8$ & 0.16 s & 0.95 & 0.72 & 0.00 & 0.00 & 0.28  & 3.73 s & 0.94 & 0.69  & 0.00 & 0.00 & 0.31 \\
$p_{1}=0.9$ & 0.17 s & 0.91 & 0.75 & 0.00 & 0.00 & 0.25 & 3.71 s & 0.90 & 0.72 & 0.00 & 0.00 & 0.28  \\
\hline
\multirow{2}{*}{$p_{0}=0.9$} & \multicolumn{6}{c}{$E(N)=2000$} & \multicolumn{6}{c}{$E(N)=10,000$}\\
\cmidrule(r){2-7} \cmidrule(r){8-13} 
& Time & $\mathcal{WA}$ & $W_{T}^{FP}$ & $W_{T}^{FN}$ & $W_{C}^{FP}$ & $W_{C}^{FN}$ & Time & $\mathcal{WA}$ & $W_{T}^{FP}$ & $W_{T}^{FN}$ & $W_{C}^{FP}$ & $W_{C}^{FN}$ \\
\hline
$p_{1}=0.2$ & 0.21 s & 0.74 & 0.00  & 0.46  & 0.54 & 0.00  & 4.15 s & 0.74 & 0.00 & 0.47 & 0.53 & 0.00 \\
$p_{1}=0.4$ & 0.19 s & 0.83 & 0.00 & 0.37 & 0.63  & 0.00 & 3.79 s & 0.82 & 0.00 & 0.38 & 0.62 & 0.00  \\
$p_{1}=0.6$ & 0.17 s & 0.91 & 0.00 & 0.25 & 0.75 & 0.00 & 3.70 s & 0.90 & 0.00 & 0.28 & 0.72 & 0.00 \\
\hline
\multicolumn{13}{c}{\textbf{Type II Randomization Designs (Simulation Scenario 2)}} \\
\hline
\multirow{2}{*}{$p_{0}=0.3$} & \multicolumn{6}{c}{$E(N)=2000$} & \multicolumn{6}{c}{$E(N)=10,000$}\\
\cmidrule(r){2-7} \cmidrule(r){8-13} 
& Time & $\mathcal{WA}$ & $W_{T}^{FP}$ & $W_{T}^{FN}$ & $W_{C}^{FP}$ & $W_{C}^{FN}$ & Time & $\mathcal{WA}$ & $W_{T}^{FP}$ & $W_{T}^{FN}$ & $W_{C}^{FP}$ & $W_{C}^{FN}$ \\
\hline
$p_{1}=0.4$ & 2.75 s & 0.99 & 0.13 & 0.00 & 0.00 & 0.87 & 6.33 s & 0.99 & 0.27 & 0.00 & 0.00 & 0.73 \\
$p_{1}=0.6$ & 2.68 s & 0.96 & 0.44 & 0.00 & 0.00 & 0.56 & 8.00 s & 0.95 & 0.44 & 0.00 & 0.00 & 0.56 \\
$p_{1}=0.8$ & 1.95 s & 0.92 & 0.54 & 0.00 & 0.00 & 0.46 & 8.68 s & 0.91 & 0.54 & 0.00 & 0.00 & 0.46 \\
\hline
\multirow{2}{*}{$p_{0}=0.6$} & \multicolumn{6}{c}{$E(N)=2000$} & \multicolumn{6}{c}{$E(N)=10,000$}\\
\cmidrule(r){2-7} \cmidrule(r){8-13} 
& Time & $\mathcal{WA}$ & $W_{T}^{FP}$ & $W_{T}^{FN}$ & $W_{C}^{FP}$ & $W_{C}^{FN}$ & Time & $\mathcal{WA}$ & $W_{T}^{FP}$ & $W_{T}^{FN}$ & $W_{C}^{FP}$ & $W_{C}^{FN}$ \\
\hline
$p_{1}=0.7$ & 2.80 s & 0.99 & 0.85 & 0.00 & 0.00 & 0.15 & 6.43 s & 0.99 & 0.73 & 0.00 & 0.00 & 0.27 \\
$p_{1}=0.8$ & 2.22 s & 0.97 & 0.76 & 0.00 & 0.00 & 0.24 & 6.19 s & 0.97 & 0.72 & 0.00 & 0.00 & 0.28 \\
$p_{1}=0.9$ & 1.45 s & 0.96 & 0.75 & 0.00 & 0.00  & 0.25 & 5.40 s & 0.95 & 0.75 & 0.00 & 0.00 & 0.25 \\
\hline
\multirow{2}{*}{$p_{0}=0.9$} & \multicolumn{6}{c}{$E(N)=2000$} & \multicolumn{6}{c}{$E(N)=10,000$}\\
\cmidrule(r){2-7} \cmidrule(r){8-13} 
& Time & $\mathcal{WA}$ & $W_{T}^{FP}$ & $W_{T}^{FN}$ & $W_{C}^{FP}$ & $W_{C}^{FN}$ & Time & $\mathcal{WA}$ & $W_{T}^{FP}$ & $W_{T}^{FN}$ & $W_{C}^{FP}$ & $W_{C}^{FN}$ \\
\hline
$p_{1}=0.2$ & 1.09 s & 0.88 & 0.00 & 0.47 & 0.53 & 0.00 & 6.86 s & 0.87 & 0.00 & 0.47 & 0.53 & 0.00 \\
$p_{1}=0.4$ & 1.45 s & 0.92 & 0.00 & 0.38 & 0.62 & 0.00 & 7.33 s & 0.91 & 0.00 & 0.39 & 0.61 & 0.00 \\
$p_{1}=0.6$ & 1.44 s & 0.96 & 0.00 & 0.25 & 0.75 & 0.00 & 5.41 s & 0.95 & 0.00 & 0.26 & 0.74 & 0.00 \\
\hline
\end{tabular}}
\end{table}

For both Simulation Scenarios 1 and 2, in each independent simulation run, we generate $\mathbf{Z}$ and $\mathbf{Y}^{*}$ according to the following procedure: 1) Given the generated $n_{i}$ and $m_{i}$, the treatments are randomly assigned within each stratum, i.e., the randomization assumption (\ref{eqn: randomization assumption}) holds. 2) We then independently generate a measured outcome $Y_{ij}^{*}$ for each study subject $ij$ according to: $Y_{ij}^{*}\sim \text{Bernoulli($p_{1}$)}$ if $Z_{ij}=1$ and $Y_{ij}^{*} \sim \text{Bernoulli($p_{0}$)}$ if $Z_{ij}=0$ (see Remark~\ref{rem: more details on simulations} for more details). We here consider testing Fisher's sharp null $H_{0}^{\text{sharp}}$. The parallel simulation studies for Neyman's weak null $H_{0}^{\text{weak}}$ are reported in Web Appendix D. After conducting 1000 independent simulation runs for each of the different prespecified sets of $(E(N), p_{0}, p_{1})$ (18 sets in total) under Simulation Scenarios 1 and 2, we report the corresponding average computation time, average warning accuracy and average sensitivity weights in Table~\ref{tab: simulation for sharp null}.

\begin{Remark}\label{rem: more details on simulations}
    Here are some further details about the specific procedure for obtaining the results in Table~\ref{tab: simulation for sharp null}: (i) As mentioned in Section~\ref{subsec: original problem} of the main text, conducting a sensitivity analysis typically makes more sense when a treatment effect is detected in a primary analysis (\citealp{rosenbaum2002observational}) based on measured outcomes $\mathbf{Y}^{*}$. Therefore, we exclude a small percent of simulation runs (i.e., generated data sets) in which the null hypothesis failed to be rejected based on the generated measured outcomes (107 out of 36,000 runs). (ii) For the remaining 35,893 simulation runs, to avoid our simulation study failing to be finished in a tolerable amount of time, we force a simulation run to stop if it exceeds 100 seconds before finding the exact solution, report the total number of such cases, and exclude such cases when calculating the average computation time, warning accuracy, and sensitivity weights. It turns out that such potentially computationally infeasible cases (i.e., the computation time for finding the exact solution exceeds 100 seconds) are very rare (5 out of 35,893 runs), and our framework can efficiently find the exact solution in most cases. (iii) The computation was done by the optimization solver \textsf{Gurobi} (version 9.1) and a laptop computer with a 1.6 GHz Dual-Core Intel Core i5 processor and 4 GB 1600 MHz DDR3 memory.
\end{Remark}

From the simulation results in Table~\ref{tab: simulation for sharp null}, we see that: 1) In most simulation runs, our computation strategy for calculating warning accuracy and sensitivity weights is very efficient, with the average computation time being a few seconds, even for large data sets (e.g., $E(N)=10,000$). 2) Other things being equal, warning accuracy decreases as measured effect size (i.e., the difference in $p_{0}$ and $p_{1}$) increases, which agrees with the fact that detection of a treatment effect is less sensitive to outcome misclassification if the measured effect size is larger. 3) Specific values of sensitivity weights should depend on the specific randomization design, sample size, and $(p_{0}, p_{1})$, but we can still observe some general patterns from the simulations. First, when a positive treatment effect was detected (i.e., $p_{1}>p_{0}$ and $H_{0}^{\text{sharp}}$ was rejected), only $W_{T}^{FP}$ and $W_{C}^{FN}$ can be non-zero. This agrees with the simple fact that there are only two types of outcome misclassification that can overturn a detected positive treatment effect: false positives among the treated group and false negatives among the control group. Similarly, when a negative treatment effect was detected (i.e., $p_{1}<p_{0}$ and $H_{0}^{\text{sharp}}$ was rejected), only $W_{T}^{FN}$ and $W_{C}^{FP}$ can be non-zero. Second, among the two non-zero sensitivity weights, which one dominates the other depends on the specific randomization design and $(p_{0}, p_{1})$. For example, for both Simulation Scenarios 1 and 2, when $p_{0}=0.3$ and $p_{1}=0.4$, $W_{C}^{FN}$ dominates $W_{T}^{FP}$, while when $p_{0}=0.6$ and $p_{1}=0.9$, $W_{T}^{FP}$ dominates $W_{C}^{FN}$ instead. In some cases, the two non-zero sensitivity weights are comparable; see $p_{0}=0.9$ and $p_{1}=0.2$ for Simulation Scenarios 1 and 2.

\vspace{-0.5 cm}

\section{Application: understanding the puzzle in the PCPT using the proposed sensitivity analysis approach}\label{sec: PCPT analysis}

\vspace{-0.2 cm}

We now apply our newly developed sensitivity analysis approach to help understand the puzzle in the PCPT (a type I randomization design) described in Section~\ref{subsubsec: PCPT}. Following \citet{thompson2003influence}, the prespecified significance level $\alpha$ is 0.05. We apply the adaptive reformulation strategy developed in Section~\ref{sec: computing} to calculate warning accuracy (equivalently, minimal alteration number) and sensitivity weights for the two binary outcomes of interest: incident prostate cancer (cancer versus no cancer) and incident high-grade prostate cancer (high-grade prostate cancer versus no cancer or low-grade cancer) at 7 years. See Table~\ref{tab: normal and severe cancer} for the results of sensitivity analyses.

\begin{table}[b]
\caption{The $p$-values, warning accuracy, and sensitivity weights for the two binary outcomes of interest in the PCPT under Fisher's sharp null and significance level $\alpha=$ 0.05.}
\label{tab: normal and severe cancer}
\small
\centering
\begin{tabular}{c|cc|cc} 
  \hline
   Outcome & \multicolumn{2}{c|}{Prostate Cancer}  & \multicolumn{2}{c}{High-Grade Prostate Cancer}\\
  (Sample Size N) & \multicolumn{2}{c|}{($N=9060$)}  & \multicolumn{2}{c}{($N=9037$)}\\
   \hline
   Relative Risk & \multicolumn{2}{c|}{ 0.75 (Protective Factor)}  & \multicolumn{2}{c}{1.27 (Risk Factor)} \\
   \hline
   $p$-value & \multicolumn{2}{c|}{$4.66 \times 10^{-13}$}  & \multicolumn{2}{c}{$6.79 \times 10^{-3}$}  \\
   \hline
  Reject $H_{0}^{\text{sharp}}$ or Not & \multicolumn{2}{c|}{Yes}  & \multicolumn{2}{c}{Yes} \\
   \hline
   Causal Conclusion & \multicolumn{2}{c|}{Prevents Prostate Cancer} & \multicolumn{2}{c}{Promotes High-Grade Cancer} \\
   \hline
Warning Accuracy & \multicolumn{2}{c|}{98.37\%}  & \multicolumn{2}{c}{99.88\%} \\
   \hline
    Minimal Alteration \# & \multicolumn{2}{c|}{147}  & \multicolumn{2}{c}{11} \\
   \hline
Sensitivity Weights  & False Positives & False Negatives  & False Positives & False Negatives  \\
   \hline
   Finasteride  & 0 & 132/147   & 2/11 & 0  \\
   Placebo & 15/147 & 0   & 0 & 9/11  \\
 \hline
 \end{tabular}
\end{table}

We now give an interpretation of the results in Table~\ref{tab: normal and severe cancer}. First, according to the reported values of warning accuracy and minimal alteration number, in a worst-case scenario sensitivity analysis, the causal conclusion concerning the prevention effect of finasteride on prostate cancer is less sensitive to outcome misclassification than the causal conclusion concerning the promotion effect on high-grade prostate cancer (the two values of warning accuracy differ by 1.5\%). A 1.5\% difference in warning accuracy is nontrivial as the sample size is large (over 9000 study subjects), and it corresponds to a difference of 136 in the minimal alteration number. In other words, to alter the causal conclusion concerning the prevention effect, it requires 147/11 $\approx$ 13.4 times more cases of outcome misclassification than that required by the causal conclusion concerning the promotion effect.

Second, we leverage the reported sensitivity weights and related prior information and expert knowledge concerning prostate cancer diagnosis to further investigate sensitivity to outcome misclassification for each of the two causal conclusions. From the reported sensitivity weights in Table~\ref{tab: normal and severe cancer}, we learn that for both of the causal conclusions, the major concern is false negatives: the dominant term among the sensitivity weights for the prevention effect is false negatives among the finasteride (treated) group and that for the promotion effect is false negatives among the placebo (control) group. For each of the two causal conclusions, is it plausible that the dominant term among the sensitivity weights is exactly the dominant term among the four actual numbers of outcome misclassification cases? We now use related prior information and expert knowledge to shed light on this issue for the two outcomes of interest. We define the following notation: $N_{T}$--number of treated subjects; $N_{C}$--number of control subjects; $p_{T,1}$ (or $p_{T,0}$)--proportion of positive (or negative) true outcomes among the treated subjects; $p_{C,1}$ (or $p_{C,0}$)--proportion of positive (or negative) true outcomes among the control subjects; $\pi_{T, 1|0}$ (or $\pi_{T, 0|1}$)--proportion of false positives (or false negatives) among the treated subjects with true outcomes being negative (or positive); $\pi_{C, 1|0}$ (or $\pi_{C, 0|1}$)--proportion of false positives (or false negatives) among the control subjects with true outcomes being negative (or positive). Then the total number of each of the four types of outcome misclassification (false positives/negatives among the treated/control group) can be decomposed into the product of these three terms, as shown in Table~\ref{tab: decomposition}. 
\begin{table}[b]
\normalsize
\caption{Decomposition of the total number of outcome misclassification cases for each of the four types of outcome misclassification.}\label{tab: decomposition}
    \centering
    \begin{tabular}{c|cc} 
  \hline
  Misclassification Cases & False Positives & False Negatives   \\
  \hline
  Finasteride (Treated)   & $N_{T}\cdot p_{T, 0}\cdot \pi_{T, 1|0}$ & $N_{T}\cdot p_{T, 1}\cdot \pi_{T, 0|1}$   \\
  Placebo (Control) & $N_{C}\cdot p_{C, 0}\cdot \pi_{C, 1|0}$ & $N_{C}\cdot p_{C, 1}\cdot \pi_{C, 0|1}$  \\
 \hline
 \end{tabular}
\end{table}

For the PCPT, we have the following related prior information or expert knowledge: (i) We have $N_{T}\approx N_{C}$ by design. (ii) Even if there are misclassified outcomes, it may still be sensible to get some sense of the values of $(p_{T, 0}, p_{T, 1}, p_{C, 0}, p_{C, 1})$ using measured outcomes. Based on measured outcomes, for the prostate cancer (all grades) outcome, $p_{T, 0}\approx 82\%, p_{T, 1}\approx 18\%, p_{C, 0}\approx 76\%, p_{C, 1}\approx 24\%$. For the high-grade prostate cancer outcome, $p_{T, 0}\approx 94\%, p_{T, 1}\approx 6\%, p_{C, 0}\approx 95\%, p_{C, 1}\approx 5\%$. Here ``$\approx$" means a very rough estimation based on measured outcomes (misclassification bias may exist). (iii) Finasteride substantially improves the detection of prostate cancer (all-grades). According to \citet{nih}: ``Finasteride has several effects on the prostate that allow better detection of prostate cancers. The drug shrinks the prostate, reducing its size and volume and increasing the chance that a biopsy will find existing cancers." Meanwhile, finasteride also greatly improves accuracy in prostate cancer grading at biopsy (\citealp{redman2008finasteride}). Therefore, we expect $\pi_{C, 0|1}\gg \pi_{T, 0|1}$ for both outcomes. 

Therefore, according to (i), (ii), and (iii), we expect that: 1) For the prostate cancer outcome, false negatives among the treated (finasteride) group (i.e., the dominant term among the four sensitivity weights) should not be the actual dominant term among the four types of outcome misclassification, as at least we expect that the number of false negatives among the treated should not dominate the number of false negatives among controls. Consequently, we expect that for the prostate cancer outcome, the actual accuracy needed to overturn the causal conclusion about the prevention effect should be lower than the warning accuracy of 98.37\%. 2) For the high-grade prostate cancer outcome, false negatives among the control (placebo) group (i.e., the dominant term among the four sensitivity weights) could be the actual dominant term among the four types of outcome misclassification, as we at least expect that the number of false negatives among controls should dominate the number of false negatives among the treated. Although whether the number of false negatives among controls dominates the number of false positives among the treated/control group needs further investigation and information, unlike the prostate cancer outcome (prevention effect), for the high-grade prostate cancer outcome we currently do not have evidence that the actual accuracy needed to overturn the causal conclusion about the promotion effect should be lower than the reported warning accuracy of 99.88\%.
 
In conclusion, based on the experimental data, related domain knowledge, and our sensitivity analysis approach for the randomization test subject to outcome misclassification, we have evidence that the causal conclusion that finasteride prevents prostate cancer should be more reliable and the conclusion that finasteride promotes high-grade prostate cancer may be due to differential outcome misclassification (i.e., $\pi_{C, 0|1}\gg \pi_{T, 0|1}$) and requires further investigations (e.g., conducting a validation study for the potential outcome misclassification bias). Our finding supports some previous arguments proposed by domain scientists and biostatisticians (e.g., \citealp{lucia2007finasteride, redman2008finasteride, shepherd2008does}). For further validation studies, the sensitivity weights reported in Table~\ref{tab: normal and severe cancer} suggest that for examining the prevention effect, priorities should be given to examining false negatives among the finasteride group, and for examining the promotion effect, more attention should be paid to false negatives among the placebo group. Also, further validation studies should be more efficient if sampling priority is given to subjects that belong to a sensitive set (see Web Appendix C for details). Additional clarifications and discussions for data application in general settings can be found in Appendix E.

\vspace{-0.5 cm}

\section{Discussion}

\vspace{-0.2 cm}

In this work, we developed a model-free and finite-population sensitivity analysis approach for randomization tests subject to outcome misclassification via large-scale integer programming. The strength of our approach is that it does not require any additional assumptions and, meanwhile, can provide useful information concerning sensitivity to outcome misclassification to help researchers analyze a randomization test subject to outcome misclassification more comprehensively and rigorously. Our strategy (i.e., adaptive integer program formulation with respect to the randomization design) for solving the ``curse of symmetry" encountered in implementing the proposed sensitivity analysis approach could shed light on other computationally intensive problems in causal inference with binary outcomes. As a real-data application, we used our sensitivity analysis approach (combined with related domain knowledge) to bring new insights into a well-known puzzle concerning the prevention and promotion effects of finasteride on prostate cancer development detected in the PCPT. A corresponding open-source \textsf{R} package \textsf{RIOM} for implementing the proposed sensitivity analysis approach has been developed and published on \textsf{GitHub} for practical and reproducible research.

As emphasized in Section~\ref{subsec: contributions}, existing model-based approaches to outcome misclassification are still very useful and can be combined with our framework (e.g., researchers can first adopt our model-free sensitivity analysis framework and then impose additional assumptions to conduct a further model-assisted secondary analysis), especially when researchers plan to leverage additional information and domain knowledge to investigate the questions beyond those studied in our framework (i.e., \textbf{Q1-Q3} in Section~\ref{subsec: contributions}) or want to provide more information and insights for \textbf{Q1-Q3}. The motivation of our framework is: what useful information concerning outcome misclassification can we learn from the experimental data before making any modeling or distributional assumptions? Although such information may not be able to address all the issues concerning outcome misclassification in randomization tests, it is immune to assumption violations and model misspecification and thus robustly maintains statistical validity. 

This work also suggests some future research directions. For example, this work focuses on randomization-based hypothesis testing subject to outcome misclassification, which is typically regarded as the first step in a randomization-based cause-and-effect analysis \citep{rosenbaum2002observational, imbens2015causal}. It is a meaningful research direction to study how to generalize the sensitivity analysis methods developed in this work to estimation and confidence interval construction for causal effects (i.e., randomization-based estimation and inference). For example, for the average treatment effect (ATE) with binary outcomes subject to outcome misclassification, a sensible strategy is to investigate how to solve the worst-case (i.e., the longest) confidence interval over all possible realizations of true outcomes $\mathbf{Y}$ subject to the constraint $\mathcal{A}(\mathbf{Y}^{*}\mid \mathbf{Y})\geq q$, where $q$ is a prespecified sensitivity parameter (i.e., the lower bounding accuracy). When $q=1$ (i.e., no outcome misclassification), the worst-case confidence interval reduces to the conventional confidence interval for the ATE without outcome misclassification. As the sensitivity parameter $q$ departs from $1$, the worst-case confidence interval for the ATE will become longer, informing about the degree of possible changes in uncertainty quantification for ATE as the accuracy of outcome classification changes. 

\vspace{-0.5 cm}

\section*{Supplementary Materials}

The appendices contain more technical details and additional simulations and discussions. Our methods have been implemented in the \textsf{R} package \textsf{RIOM}, which is publicly available at \url{https://github.com/siyuheng/RIOM}.

\vspace{-0.5 cm}

\section*{Acknowledgements}

The authors thank the PCPT investigators and the Southwest Oncology Group (SWOG) for the use of the PCPT data. The authors thank Professor Dylan Small for his helpful advice and comments.

\section*{Funding}

The work of Siyu Heng was supported by NIH Grants R01AI131771 and R21DA060433, a grant from the New York University Research Catalyst Prize, and a New York University School of Global Public Health Research Support Grant. The work of Pamela Shaw was supported by NIH Grant R01AI131771. 

\vspace{-0.5 cm}

%% if your bibliography is in bibtex format, uncomment commands:
\bibliographystyle{apalike} % Style BST file
\bibliography{ref}       % Bibliography file (usually '*.bib')

\newpage

\begin{center}
    \Large \bf Appendices for ``Sensitivity Analysis for Binary Outcome Misclassification in Randomization Tests via Integer Programming"
\end{center}

\begingroup
\allowdisplaybreaks

\section*{Web Appendix A: Review and Some Preliminary Results}

\subsection*{A.1: Review of Neyman's weak null hypothesis and the corresponding test statistic}

As mentioned in the main text, in addition to Fisher's sharp null $H_{0}^{\text{sharp}}$, another extensively considered null hypothesis is Neyman's weak null of no average treatment effect $H_{0}^{\text{weak}}: \frac{1}{N} \sum_{i=1}^{I}\sum_{j=1}^{n_{i}}(Y_{ij}(1)-Y_{ij}(0))=0$ (\citealp{neyman1923application}), which can be rewritten as $H_{0}^{\text{weak}}: \sum_{i=1}^{I}\frac{n_{i}}{N} \tau_{i}=0$, where $\tau_{i}=\frac{1}{n_{i}}\sum_{j=1}^{n_{i}}Y_{ij}(1)-\frac{1}{n_{i}}\sum_{j=1}^{n_{i}}Y_{ij}(0)$. A commonly used randomization test statistic for testing $H_{0}^{\text{weak}}$ is the (randomization-based) difference-in-means estimator (i.e., the Neyman estimator): $T_{\text{Neyman}}(\mathbf{Z}, \mathbf{Y})=\sum_{i=1}^{I}\frac{n_{i}}{N} \hat{\tau}_{i}$, where $\hat{\tau}_{i}=\frac{1}{m_{i}}\sum_{j=1}^{n_{i}}Z_{ij}Y_{ij}-\frac{1}{n_{i}- m_{i}}\sum_{j=1}^{n_{i}}(1-Z_{ij})Y_{ij}$ (\citealp{neyman1923application}). Under $H_{0}^{\text{weak}}$, we have $E(T_{\text{Neyman}})=0$ and $\text{Var}(T_{\text{Neyman}})=\sum_{i=1}^{I} (\frac{n_{i}}{N})^{2}\text{Var}(\hat{\tau}_{i})$, where 
    $\text{Var}(\hat{\tau}_{i})=\frac{S_{T, i}^{2}}{m_{i}}+\frac{S_{C, i}^{2}}{n_{i}-m_{i}}-\frac{S_{i}^{2}}{n_{i}}$, with $S_{T, i}^{2}=\frac{1}{n_{i}-1}\sum_{j=1}^{n_{i}}\big(Y_{ij}(1)-\frac{1}{n_{i}}\sum_{j=1}^{n_{i}}Y_{ij}(1)\big)^{2}$, $S_{C, i}^{2}=\frac{1}{n_{i}-1}\sum_{j=1}^{n_{i}}\big(Y_{ij}(0)-\frac{1}{n_{i}}\sum_{j=1}^{n_{i}}Y_{ij}(0)\big)^{2}$, and $S_{i}^{2}=\frac{1}{n_{i}-1}\sum_{j=1}^{n_{i}}\big(Y_{ij}(1)-Y_{ij}(0)-\tau_{i} \big)^{2}$. Researchers commonly use the following finite-population central limit theorem (\citealp{imbens2015causal, li2017general}) to test $H_{0}^{\text{weak}}$: $\frac{T_{\text{Neyman}}}{\sqrt{\text{Var}(T_{\text{Neyman}})}} \xrightarrow{\mathcal{L}} N(0,1)$ under $H_{0}^{\text{weak}}$. Since $S_{i}^{2}$ involves $\tau_{i}$, which cannot be identified from the observed data, researchers commonly adopt the Neyman variance estimator $\widehat{\text{Var}}(T_{\text{Neyman}})$ for $\text{Var}(T_{\text{Neyman}})$ (\citealp{neyman1923application}): $\widehat{\text{Var}}(T_{\text{Neyman}})=\sum_{i=1}^{I} (\frac{n_{i}}{N})^{2}\cdot (\frac{\widehat{S}_{T, i}^{2}}{m_{i}}+\frac{\widehat{S}_{C, i}^{2}}{n_{i}-m_{i}})$, where $\widehat{S}_{T, i}^{2}=\frac{1}{m_{i}-1}\sum_{j=1}^{n_{i}}Z_{ij}(Y_{ij}-\frac{1}{m_{i}}\sum_{j=1}^{n_{i}}Z_{ij}Y_{ij})^{2}$ and $\widehat{S}_{C, i}^{2}=\frac{1}{n_{i}-m_{i}-1}\sum_{j=1}^{n_{i}}(1-Z_{ij})\{Y_{ij}-\frac{1}{n_{i}-m_{i}}\sum_{j=1}^{n_{i}}(1-Z_{ij})Y_{ij}\}^{2}$ (if $\min \{ m_{i}, n_{i}-m_{i}\} \neq 1$). For example, in a two-sided level-$\alpha$ test adopting $\widehat{\text{Var}}(T_{\text{Neyman}})$, researchers reject $H_{0}^{\text{weak}}$ if and only if $\frac{T_{\text{Neyman}}^{2}}{\widehat{\text{Var}}(T_{\text{Neyman}})}> \chi^{2}_{1, 1-\alpha}$ (\citealp{imbens2015causal}). If $\min\{m_{i}, n_{i}\}=1$, we can adopt the method in \citet{fogarty2017randomization} to estimate $\text{Var}(T_{\text{Neyman}})$.

\subsection*{A.2: Some preliminary results concerning computing warning accuracy with testing Fisher's sharp null}

In integer programming literature, the general standard form of an integer quadratically constrained linear program (IQCLP) (\citealp{lee2011mixed, burer2012milp, conforti2014integer}) is as below:
\begin{equation*}
     \begin{split}
        \underset{\mathbf{x}}{\text{maximize}} \quad & \mathbf{q}^T \mathbf{x} +c ~~~\text{(linear objective function)} \\
         \text{subject to}\quad & \mathbf{x}^T \mathbf{Q}_{k} \mathbf{x} + \mathbf{q}_{k}^T \mathbf{x} \leq b_{k}, \forall k ~~~\text{(quadratic constraints)} \\
          & \mathbf{A}\mathbf{x} \leq \mathbf{b},~~~\text{(linear constraints)} \\
         & \mathbf{l} \leq \mathbf{x} \leq \mathbf{u}, ~~~\text{(box constraints)} \\
        & \text{Some or all elements of} ~\mathbf{x}~ \text{are integers.}~~~\text{(integer constraints)}
     \end{split}
 \end{equation*}

In this section, we give some preliminary results for writing the integer programs (P0), (P1), and (P2) in the above standard form. Specifically, we would like to rewrite every quadratic term appearing in some quadratic constraint in each integer program considered in this paper as the standard form $\mathbf{x}^T \mathbf{Q} \mathbf{x} + \mathbf{q}^T \mathbf{x}$, which is required by optimization solvers such as \textsf{Gurobi} (\citealp{gurobi}). For the quadratic constraint in the integer program (P0), we have
 \begin{align*}
   & [T_{\text{M-H}}(\mathbf{Z}, \mathbf{Y})-E\{T_{\text{M-H}}(\mathbf{Z}, \mathbf{Y})\}]^{2}-\chi^{2}_{1, 1-\alpha} \cdot \text{Var}\{T_{\text{M-H}}(\mathbf{Z}, \mathbf{Y})\}\\
   &=\Big\{\sum_{i=1}^{I}\sum_{j=1}^{n_{i}}Z_{ij}Y_{ij}-\sum_{i=1}^{I}\Big(\frac{m_{i}}{n_{i}}\sum_{j=1}^{n_{i}}Y_{ij}\Big)\Big\}^{2}-\chi^{2}_{1, 1-\alpha} \cdot\sum_{i=1}^{I}\frac{m_{i}(\sum_{j=1}^{n_{i}}Y_{ij})(n_{i}-\sum_{j=1}^{n_{i}}Y_{ij})(n_{i}-m_{i})}{n_{i}^{2}(n_{i}-1)}\\
   &=\sum_{i=1}^{I}\sum_{j=1}^{n_{i}}\Big\{\Big(Z_{ij}-\frac{m_{i}}{n_{i}}\Big)^{2}+\chi^{2}_{1, 1-\alpha} \cdot \frac{m_{i}(n_{i}-m_{i})}{n_{i}^{2}(n_{i}-1)}\Big\}Y_{ij}^{2}\\
   &\quad \quad  +\sum_{i=1}^{I}\sum_{j\neq j^{\prime}}\Big\{\Big(Z_{ij}-\frac{m_{i}}{n_{i}}\Big)\Big(Z_{ij^{\prime}} -\frac{m_{i }}{n_{i}}\Big)+\chi^{2}_{1, 1-\alpha} \cdot \frac{m_{i}(n_{i}-m_{i})}{n_{i}^{2}(n_{i}-1)} \Big\}Y_{ij}Y_{ij^{\prime}} \\
   &\quad \quad +\sum_{i\neq i^{\prime}}\sum_{j=1}^{n_{i}}\sum_{j^{\prime}=1}^{n_{i^{\prime}}} \Big(Z_{ij}-\frac{m_{i}}{n_{i}}\Big)\Big(Z_{i^{\prime}j^{\prime}} -\frac{m_{i^{\prime} }}{n_{i^{\prime}}}\Big)Y_{ij}Y_{i^{\prime}j^{\prime}}-\chi^{2}_{1, 1-\alpha} \cdot\sum_{i=1}^{I}\sum_{j=1}^{n_{i}}\frac{m_{i}n_{i}(n_{i}-m_{i})}{n_{i}^{2}(n_{i}-1)}Y_{ij}.
\end{align*}  
The above equation can be rewritten as
\begin{align*}
   & [T_{\text{M-H}}(\mathbf{Z}, \mathbf{Y})-E\{T_{\text{M-H}}(\mathbf{Z}, \mathbf{Y})\}]^{2}-\chi^{2}_{1, 1-\alpha} \cdot \text{Var}\{T_{\text{M-H}}(\mathbf{Z}, \mathbf{Y})\}\\
   &=\Big\{\sum_{i=1}^{I}\sum_{j=1}^{n_{i}}Z_{ij}Y_{ij}-\sum_{i=1}^{I}\Big(\frac{m_{i}}{n_{i}}\sum_{j=1}^{n_{i}}Y_{ij}\Big)\Big\}^{2}-\chi^{2}_{1, 1-\alpha} \cdot\sum_{i=1}^{I}\frac{m_{i}(\sum_{j=1}^{n_{i}}Y_{ij})(n_{i}-\sum_{j=1}^{n_{i}}Y_{ij})(n_{i}-m_{i})}{n_{i}^{2}(n_{i}-1)}\\
   &=\Big[ \sum_{i=1}^{I}(\Upsilon_{i}^{10}+\Upsilon_{i}^{11})-\sum_{i=1}^{I}\Big\{\frac{m_{i}}{n_{i}}(\Upsilon_{i}^{00}+\Upsilon_{i}^{01}+\Upsilon_{i}^{10}+\Upsilon_{i}^{11}) \Big\} \Big]^{2}\\
   &\quad \quad \quad - \chi^{2}_{1, 1-\alpha} \cdot \sum_{i=1}^{I}\frac{m_{i}(\Upsilon_{i}^{00}+\Upsilon_{i}^{01}+\Upsilon_{i}^{10}+\Upsilon_{i}^{11})(n_{i}-\Upsilon_{i}^{00}-\Upsilon_{i}^{01}-\Upsilon_{i}^{10}-\Upsilon_{i}^{11})(n_{i}-m_{i})}{n_{i}^{2}(n_{i}-1)}\\
   &=\sum_{i=1}^{I}\Big[ \Big\{ \frac{m_{i}^{2}}{n_{i}^{2}}+\chi^{2}_{1, 1-\alpha}\cdot \frac{m_{i}(n_{i}-m_{i}) }{n_{i}^{2}(n_{i}-1)}  \Big\}\Big(\Upsilon_{i}^{00}\Big)^{2}\\
   &\quad \quad \quad \quad \quad \quad \quad  +\Big\{ \frac{m_{i}^{2}}{n_{i}^{2}}+\chi^{2}_{1, 1-\alpha}\cdot \frac{m_{i}(n_{i}-m_{i})}{n_{i}^{2}(n_{i}-1)}  \Big\}\Big(\Upsilon_{i}^{01}\Big)^{2}\\
   &\quad \quad \quad \quad \quad \quad \quad  +\Big\{ \Big(1-\frac{m_{i}}{n_{i}}\Big)^{2}+\chi^{2}_{1, 1-\alpha}\cdot \frac{m_{i}(n_{i}-m_{i})}{n_{i}^{2}(n_{i}-1)}  \Big\}\Big(\Upsilon_{i}^{10}\Big)^{2}\\
   &\quad \quad \quad \quad \quad \quad \quad +\Big\{ \Big(1-\frac{m_{i}}{n_{i}}\Big)^{2}+\chi^{2}_{1, 1-\alpha}\cdot \frac{m_{i}(n_{i}-m_{i})}{n_{i}^{2}(n_{i}-1)}  \Big\}\Big(\Upsilon_{i}^{11}\Big)^{2}\\
    &\quad \quad \quad \quad \quad \quad \quad +2\Big\{ \frac{m_{i}^{2}}{n_{i}^{2}}+\chi^{2}_{1, 1-\alpha}\cdot \frac{m_{i}(n_{i}-m_{i}) }{n_{i}^{2}(n_{i}-1)}  \Big\}\Upsilon_{i}^{00}\Upsilon_{i}^{01}\\
    &\quad \quad \quad \quad \quad \quad \quad +2\Big\{- \frac{m_{i}}{n_{i}}\Big(1-\frac{m_{i}}{n_{i}}\Big)+\chi^{2}_{1, 1-\alpha}\cdot \frac{m_{i}(n_{i}-m_{i})}{n_{i}^{2}(n_{i}-1)} \Big\}\Upsilon_{i}^{00}\Upsilon_{i}^{10}\\
    &\quad \quad \quad \quad \quad \quad \quad +2\Big\{- \frac{m_{i}}{n_{i}}\Big(1-\frac{m_{i}}{n_{i}}\Big)+\chi^{2}_{1, 1-\alpha}\cdot \frac{m_{i}(n_{i}-m_{i})}{n_{i}^{2}(n_{i}-1)} \Big\}\Upsilon_{i}^{00}\Upsilon_{i}^{11}\\
    &\quad \quad \quad \quad \quad \quad \quad +2\Big\{- \frac{m_{i}}{n_{i}}\Big(1-\frac{m_{i}}{n_{i}}\Big)+\chi^{2}_{1, 1-\alpha}\cdot \frac{m_{i}(n_{i}-m_{i})}{n_{i}^{2}(n_{i}-1)} \Big\}\Upsilon_{i}^{01}\Upsilon_{i}^{10}\\
    &\quad \quad \quad \quad \quad \quad \quad +2\Big\{- \frac{m_{i}}{n_{i}}\Big(1-\frac{m_{i}}{n_{i}}\Big)+\chi^{2}_{1, 1-\alpha}\cdot \frac{m_{i}(n_{i}-m_{i})}{n_{i}^{2}(n_{i}-1)} \Big\}\Upsilon_{i}^{01}\Upsilon_{i}^{11}\\
    &\quad \quad \quad \quad \quad \quad \quad +2\Big\{ \Big(1-\frac{m_{i}}{n_{i}}\Big)^{2}+\chi^{2}_{1, 1-\alpha}\cdot \frac{m_{i}(n_{i}-m_{i})}{n_{i}^{2}(n_{i}-1)}  \Big\}\Upsilon_{i}^{10}\Upsilon_{i}^{11}\\
    &\quad \quad \quad \quad \quad \quad \quad -\chi^{2}_{1, 1-\alpha}\cdot \frac{m_{i}n_{i}(n_{i}-m_{i})}{n_{i}^{2}(n_{i}-1)}\Big(\Upsilon_{i}^{00}+\Upsilon_{i}^{01}+\Upsilon_{i}^{10}+\Upsilon_{i}^{11}\Big) \Big]\\
    &\quad+\sum_{i\neq i^{\prime} }\Big\{\frac{m_{i}}{n_{i}}\frac{m_{i^{\prime} }}{n_{i^{\prime}}}\Upsilon_{i}^{00}\Upsilon_{i^{\prime}}^{00}+\frac{m_{i}}{n_{i}}\frac{m_{i^{\prime} }}{n_{i^{\prime}}}\Upsilon_{i}^{00}\Upsilon_{i^{\prime}}^{01}-\frac{m_{i}}{n_{i}}\Big(1-\frac{m_{i^{\prime} }}{n_{i^{\prime}}}\Big)\Upsilon_{i}^{00}\Upsilon_{i^{\prime}}^{10}\\
    &\quad \quad \quad \quad \quad \quad \quad -\frac{m_{i}}{n_{i}}\Big(1-\frac{m_{i^{\prime} }}{n_{i^{\prime}}}\Big)\Upsilon_{i}^{00}\Upsilon_{i^{\prime}}^{11}  +\frac{m_{i}}{n_{i}}\frac{m_{i^{\prime} }}{n_{i^{\prime}}}\Upsilon_{i}^{01}\Upsilon_{i^{\prime}}^{00}+\frac{m_{i}}{n_{i}}\frac{m_{i^{\prime} }}{n_{i^{\prime}}}\Upsilon_{i}^{01}\Upsilon_{i^{\prime}}^{01}\\
    &\quad \quad \quad \quad \quad \quad \quad -\frac{m_{i}}{n_{i}}\Big(1-\frac{m_{i^{\prime} }}{n_{i^{\prime}}}\Big)\Upsilon_{i}^{01}\Upsilon_{i^{\prime}}^{10}-\frac{m_{i}}{n_{i}}\Big(1-\frac{m_{i^{\prime} }}{n_{i^{\prime}}}\Big)\Upsilon_{i}^{01}\Upsilon_{i^{\prime}}^{11}-\Big(1-\frac{m_{i }}{n_{i}}\Big)\frac{m_{i^{\prime} }}{n_{i^{\prime} }}\Upsilon_{i}^{10}\Upsilon_{i^{\prime}}^{00}\\
    &\quad \quad \quad \quad \quad \quad \quad-\Big(1-\frac{m_{i }}{n_{i}}\Big)\frac{m_{i^{\prime} }}{n_{i^{\prime} }}\Upsilon_{i}^{10}\Upsilon_{i^{\prime} }^{01}+\Big(1-\frac{m_{i}}{n_{i}}\Big)\Big(1-\frac{m_{i^{\prime} }}{n_{i^{\prime}}}\Big)\Upsilon_{i}^{10}\Upsilon_{i^{\prime}}^{10}\\
    &\quad \quad \quad \quad \quad \quad \quad +\Big(1-\frac{m_{i}}{n_{i}}\Big)\Big(1-\frac{m_{i^{\prime} }}{n_{i^{\prime}}}\Big)\Upsilon_{i}^{10}\Upsilon_{i^{\prime}}^{11}-\Big(1-\frac{m_{i }}{n_{i}}\Big)\frac{m_{i^{\prime} }}{n_{i^{\prime} }}\Upsilon_{i}^{11}\Upsilon_{i^{\prime}}^{00}-\Big(1-\frac{m_{i }}{n_{i}}\Big)\frac{m_{i^{\prime} }}{n_{i^{\prime} }}\Upsilon_{i}^{11}\Upsilon_{i^{\prime} }^{01}\\
    &\quad \quad \quad \quad \quad \quad \quad +\Big(1-\frac{m_{i}}{n_{i}}\Big)\Big(1-\frac{m_{i^{\prime} }}{n_{i^{\prime}}}\Big)\Upsilon_{i}^{11}\Upsilon_{i^{\prime}}^{10}+\Big(1-\frac{m_{i}}{n_{i}}\Big)\Big(1-\frac{m_{i^{\prime} }}{n_{i^{\prime}}}\Big)\Upsilon_{i}^{11}\Upsilon_{i^{\prime}}^{11}\Big\},
\end{align*}  
which enters into the quadratic constraint for the integer program (P1), of which a standard form is given in Web Appendix B.1. The above equation can also be written as
\begin{align*}
  & [T_{\text{M-H}}(\mathbf{Z}, \mathbf{Y})-E\{T_{\text{M-H}}(\mathbf{Z}, \mathbf{Y})\}]^{2}-\chi^{2}_{1, 1-\alpha} \cdot \text{Var}\{T_{\text{M-H}}(\mathbf{Z}, \mathbf{Y})\}\\
   &=\Big[ \sum_{s=1}^{S}\sum_{p=1}^{\widetilde{N}_{s}}d_{sp}(\Delta_{sp}^{10}+\Delta_{sp}^{11})-\sum_{s=1}^{S}\sum_{p=1}^{\widetilde{N}_{s}}d_{sp}\Big\{\frac{\widetilde{m}_{s}}{\widetilde{n}_{s}}(\Delta_{sp}^{00}+\Delta_{sp}^{01}+\Delta_{sp}^{10}+\Delta_{sp}^{11}) \Big\} \Big]^{2}\\
   &\quad \quad - \chi^{2}_{1, 1-\alpha} \cdot \sum_{s=1}^{S}\sum_{p=1}^{\widetilde{N}_{s}}d_{sp}\frac{\widetilde{m}_{s}(\Delta_{sp}^{00}+\Delta_{sp}^{01}+\Delta_{sp}^{10}+\Delta_{sp}^{11})(\widetilde{n}_{s}-\Delta_{sp}^{00}-\Delta_{sp}^{01}-\Delta_{sp}^{10}-\Delta_{sp}^{11})(\widetilde{n}_{s}-\widetilde{m}_{s})}{\widetilde{n}_{s}^{2}(\widetilde{n}_{s}-1)}\\
   &=\sum_{s=1}^{S}\sum_{p=1}^{\widetilde{N}_{s}}\sum_{s^{\prime}=1}^{S}\sum_{p^{\prime}=1}^{\widetilde{N}_{s^{\prime} }}d_{sp}d_{s^{\prime}p^{\prime}} \Big[ \big \{ \Delta_{sp}^{10}+\Delta_{sp}^{11}-\frac{\widetilde{m}_{s}}{\widetilde{n}_{s}}(\Delta_{sp}^{00}+\Delta_{sp}^{01}+\Delta_{sp}^{10}+\Delta_{sp}^{11})\big \}\\
   &\quad \quad \quad \quad \quad \quad \quad \quad \quad \quad \quad \quad \quad  \times \big \{ \Delta_{s^{\prime}p^{\prime}}^{10}+\Delta_{s^{\prime}p^{\prime}}^{11}-\frac{\widetilde{m}_{s^{\prime} }}{\widetilde{n}_{s^{\prime} }}(\Delta_{s^{\prime}p^{\prime}}^{00}+\Delta_{s^{\prime}p^{\prime}}^{01}+\Delta_{s^{\prime}p^{\prime}}^{10}+\Delta_{s^{\prime}p^{\prime}}^{11})\big \}\Big]\\
   &\quad \quad  -\chi^{2}_{1, 1-\alpha} \cdot \sum_{s=1}^{S}\sum_{p=1}^{\widetilde{N}_{s}}d_{sp}\frac{\widetilde{m}_{s}(\Delta_{sp}^{00}+\Delta_{sp}^{01}+\Delta_{sp}^{10}+\Delta_{sp}^{11})(\widetilde{n}_{s}-\Delta_{sp}^{00}-\Delta_{sp}^{01}-\Delta_{sp}^{10}-\Delta_{sp}^{11})(\widetilde{n}_{s}-\widetilde{m}_{s})}{\widetilde{n}_{s}^{2}(\widetilde{n}_{s}-1)},
\end{align*}
which enters into the quadratic constraint for the integer program (P2), of which a standard form is given in Web Appendix B.2.

\subsection*{A.3: Some preliminary results concerning computing warning accuracy with Neyman's weak null}

By Definition 1 in the main text, if Neyman's weak null hypothesis $H_{0}^{\text{weak}}$ was rejected based on measured outcomes (i.e., $\{T_{\text{Neyman}}(\mathbf{Z}, \mathbf{Y}^{*})\}^{2}-\chi^{2}_{1,1-\alpha} \cdot \widehat{\text{Var}}\{T_{\text{Neyman}}(\mathbf{Z}, \mathbf{Y}^{*})\}> 0$), the warning accuracy $\mathcal{WA}$ for testing $H_{0}^{\text{weak}}$ with the Neyman estimator is the optimal value of the following integer quadratically constrained linear program (IQCLP):
\begin{equation*}
     \begin{split}
        \underset{\mathbf{Y}\in \{0,1\}^{N} } {\text{maximize}} \quad &\frac{1}{N}\sum_{i=1}^{I}\sum_{j=1}^{n_{i}}Y_{ij}^{*}Y_{ij}+\frac{1}{N}\sum_{i=1}^{I}\sum_{j=1}^{n_{i}}(1-Y_{ij}^{*})(1-Y_{ij}) \quad \quad (\text{P}0^{\prime})\\
         \text{subject to}\quad & \{T_{\text{Neyman}}(\mathbf{Z}, \mathbf{Y})\}^{2}-\chi^{2}_{1,1-\alpha} \cdot \widehat{\text{Var}}\{T_{\text{Neyman}}(\mathbf{Z}, \mathbf{Y})\} \leq 0.
     \end{split}
 \end{equation*}
 If $H_{0}^{\text{weak}}$ fails to be rejected based on measured outcomes (i.e., $\{T_{\text{Neyman}}(\mathbf{Z}, \mathbf{Y}^{*})\}^{2}-\chi^{2}_{1,1-\alpha} \cdot \widehat{\text{Var}}\{T_{\text{Neyman}}(\mathbf{Z}, \mathbf{Y}^{*})\}\leq 0$), we just need to replace the ``$\leq 0$" with the ``$\geq 0$" in the quadratic constraint in $(\text{P}0^{\prime})$. As mentioned in the main text, in this paper, we will focus on ($\text{P}0^{\prime}$). We here try to rewrite every quadratic term in the integer programs ($\text{P}0^{\prime})$, (P3), and (P4) considered in Web Appendices A and B as the standard form $\mathbf{x}^T \mathbf{Q} \mathbf{x} + \mathbf{q}^T \mathbf{x}$, which is required by optimization solvers such as \textsf{Gurobi} (\citealp{gurobi}). For the quadratic constraint in the integer program ($\text{P}0^{\prime}$), we have
\begin{align*}
    &\{T_{\text{Neyman}}(\mathbf{Z}, \mathbf{Y})\}^{2}-\chi^{2}_{1, 1-\alpha} \cdot \widehat{\text{Var}}\{T_{\text{Neyman}}(\mathbf{Z}, \mathbf{Y})\}\\
&=\sum_{i=1}^{I}\sum_{j=1}^{n_{i}}\Big[\Big\{ \frac{n_{i}}{Nm_{i}}Z_{ij}-\frac{n_{i}}{N(n_{i}-m_{i})}(1-Z_{ij})\Big\}^{2}\\
&\quad \quad \quad \quad \quad \quad  -\chi^{2}_{1, 1-\alpha}\cdot \frac{(\frac{n_{i}}{N})^{2}}{m_{i}(m_{i}-1)}Z_{ij} \big(1-\frac{Z_{ij}}{m_{i}}\big)^{2}\\
&\quad \quad \quad \quad \quad \quad -\chi^{2}_{1, 1-\alpha}\cdot \frac{(\frac{n_{i}}{N})^{2}}{m_{i}(m_{i}-1)}\sum_{j^{\prime}\neq j}\frac{Z_{ij^{\prime}} Z_{ij}^{2}}{m_{i}^{2}}\\
&\quad \quad \quad \quad \quad \quad  -\chi^{2}_{1, 1-\alpha}\cdot \frac{(\frac{n_{i}}{N})^{2}}{(n_{i}-m_{i})(n_{i}-m_{i}-1)}(1-Z_{ij}) \big(1-\frac{1-Z_{ij}}{n_{i}-m_{i}}\big)^{2}\\
&\quad \quad \quad \quad \quad \quad -\chi^{2}_{1, 1-\alpha}\cdot \frac{(\frac{n_{i}}{N})^{2}}{(n_{i}-m_{i})(n_{i}-m_{i}-1)}\sum_{j^{\prime}\neq j}\frac{(1-Z_{ij^{\prime}})(1-Z_{ij})^{2}}{(n_{i}-m_{i})^{2}}\Big] Y_{ij}^{2}\\
&\quad +\sum_{i=1}^{I}\sum_{j\neq j^{\prime}}\Big[ \Big\{ \frac{n_{i}}{Nm_{i}}Z_{ij}-\frac{n_{i}}{N(n_{i}-m_{i})}(1-Z_{ij})\Big\}\Big\{ \frac{n_{i}}{Nm_{i}}Z_{ij^{\prime} }-\frac{n_{i}}{N(n_{i}-m_{i})}(1-Z_{ij^{\prime} })\Big\}\\
&\quad \quad \quad \quad \quad \quad + \chi^{2}_{1, 1-\alpha}\cdot \frac{(\frac{n_{i}}{N})^{2}}{m_{i}(m_{i}-1)}Z_{ij}\big(1-\frac{Z_{ij}}{m_{i}}\big)\frac{Z_{ij^{\prime}}}{m_{i}}\\
&\quad \quad \quad \quad \quad \quad +\chi^{2}_{1, 1-\alpha}\cdot \frac{(\frac{n_{i}}{N})^{2}}{m_{i}(m_{i}-1)}Z_{ij^{\prime} }\big(1-\frac{Z_{ij^{\prime}}}{m_{i}}\big)\frac{Z_{ij}}{m_{i}}\\
&\quad \quad \quad \quad \quad \quad -\chi^{2}_{1, 1-\alpha}\cdot \frac{(\frac{n_{i}}{N})^{2}}{m_{i}(m_{i}-1)}\sum_{j^{\prime \prime}\neq j, j^{\prime}}\frac{Z_{ij^{\prime \prime}}Z_{ij}Z_{ij^{\prime}}}{m_{i}^{2}}\\
&\quad \quad \quad \quad \quad \quad + \chi^{2}_{1, 1-\alpha}\cdot \frac{(\frac{n_{i}}{N})^{2}}{(n_{i}-m_{i})(n_{i}-m_{i}-1)}(1-Z_{ij})\big(1-\frac{1-Z_{ij}}{n_{i}-m_{i}}\big)\frac{1-Z_{ij^{\prime}}}{n_{i}-m_{i}}\\
&\quad \quad \quad \quad \quad \quad + \chi^{2}_{1, 1-\alpha}\cdot \frac{(\frac{n_{i}}{N})^{2}}{(n_{i}-m_{i})(n_{i}-m_{i}-1)}(1-Z_{ij^{\prime}})\big(1-\frac{1-Z_{ij^{\prime}}}{n_{i}-m_{i}}\big)\frac{1-Z_{ij}}{n_{i}-m_{i}}\\
& \quad \quad \quad \quad \quad \quad - \chi^{2}_{1, 1-\alpha}\cdot \frac{(\frac{n_{i}}{N})^{2}}{(n_{i}-m_{i})(n_{i}-m_{i}-1)}\sum_{j^{\prime \prime}\neq j, j^{\prime}}\frac{(1-Z_{ij^{\prime \prime}})(1-Z_{ij})(1-Z_{ij^{\prime}})}{(n_{i}-m_{i})^{2}}\Big] Y_{ij}Y_{ij^{\prime}}\\
&\quad +\sum_{i\neq i^{\prime}}\sum_{j=1}^{n_{i}}\sum_{j^{\prime}=1}^{n_{i^{\prime}}}\Big\{ \frac{n_{i}}{Nm_{i }}Z_{ij}-\frac{n_{i}}{N(n_{i}-m_{i})}(1-Z_{ij})\Big\}\Big\{ \frac{n_{i^{\prime} }}{Nm_{i^{\prime} }}Z_{i^{\prime} j^{\prime} }-\frac{n_{i^{\prime} }}{N(n_{i^{\prime} }-m_{i^{\prime} })}(1-Z_{i^{\prime} j^{\prime} })\Big\} Y_{ij}Y_{i^{\prime} j^{\prime}}.
\end{align*}
The above equation can be rewritten as
\begin{align*}
    &\{T_{\text{Neyman}}(\mathbf{Z}, \mathbf{Y})\}^{2}-\chi^{2}_{1, 1-\alpha} \cdot \widehat{\text{Var}}\{T_{\text{Neyman}}(\mathbf{Z}, \mathbf{Y})\}\\
    &=\Big[\sum_{i=1}^{I}\frac{n_{i}}{N}\Big\{ \frac{1}{m_{i}}\sum_{j=1}^{n_{i}}Z_{ij}Y_{ij}-\frac{1}{n_{i}- m_{i}}\sum_{j=1}^{n_{i}}(1-Z_{ij})Y_{ij}\Big\}\Big]^{2}\\
    &\quad \quad  \quad -\chi^{2}_{1, 1-\alpha} \cdot \sum_{i=1}^{I} \Big(\frac{n_{i}}{N}\Big)^{2} \Big[\frac{1}{m_{i}(m_{i}-1)}\sum_{j=1}^{n_{i}}Z_{ij}\big(Y_{ij}-\frac{1}{m_{i}}\sum_{j^{\prime}=1}^{n_{i}}Z_{ij^{\prime}}Y_{ij^{\prime}}\big)^{2} \nonumber \\
&\quad \quad \quad \quad \quad +\frac{1}{(n_{i}-m_{i})(n_{i}-m_{i}-1)}\sum_{j=1}^{n_{i}}(1-Z_{ij})\Big\{Y_{ij}-\frac{1}{n_{i}-m_{i}}\sum_{j^{\prime}=1}^{n_{i}}(1-Z_{ij^{\prime}})Y_{ij^{\prime}}\Big\}^{2}\Big]\\
&=\Big[ \sum_{i=1}^{I}\Big\{\frac{n_{i}}{Nm_{i}}(\Upsilon_{i}^{10}+\Upsilon_{i}^{11}) -\frac{n_{i}}{N(n_{i}-m_{i})}(\Upsilon_{i}^{00}+\Upsilon_{i}^{01}) \Big\} \Big]^{2}\\
&\quad \quad \quad -\chi^{2}_{1, 1-\alpha} \cdot \sum_{i=1}^{I} \Big(\frac{n_{i}}{N}\Big)^{2} \Big\{ \frac{\Upsilon_{i}^{10}+\Upsilon_{i}^{11} }{m_{i}(m_{i}-1)} -\frac{(\Upsilon_{i}^{10}+\Upsilon_{i}^{11})^{2} }{m_{i}^{2}(m_{i}-1)}\\
&\quad \quad \quad \quad \quad \quad \quad \quad \quad + \frac{\Upsilon_{i}^{00}+\Upsilon_{i}^{01} }{(n_{i}-m_{i})(n_{i}-m_{i}-1)}-\frac{(\Upsilon_{i}^{00}+\Upsilon_{i}^{01})^{2} }{(n_{i}-m_{i})^{2}(n_{i}-m_{i}-1)}\Big\}\\
&= \sum_{i=1}^{I}\Big[\Big\{ \frac{n_{i}^{2}}{N^{2}(n_{i}-m_{i})^{2} }+\chi^{2}_{1, 1-\alpha} \cdot \frac{n_{i}^{2}}{N^{2}(n_{i}-m_{i})^{2}(n_{i}-m_{i}-1)}\Big\} \Big\{ (\Upsilon_{i}^{00})^{2}+(\Upsilon_{i}^{01})^{2}\Big\}\\
&\quad \quad \quad \quad \quad + \Big\{ \frac{n_{i}^{2}}{N^{2}m_{i}^{2} }+\chi^{2}_{1, 1-\alpha} \cdot \frac{n_{i}^{2}}{N^{2}m_{i}^{2}(m_{i}-1)}\Big\} \Big\{ (\Upsilon_{i}^{10})^{2}+(\Upsilon_{i}^{11})^{2}\Big\}\\
&\quad \quad \quad \quad \quad +2 \Big\{ \frac{n_{i}^{2}}{N^{2}(n_{i}-m_{i})^{2} }+\chi^{2}_{1, 1-\alpha} \cdot \frac{n_{i}^{2}}{N^{2}(n_{i}-m_{i})^{2}(n_{i}-m_{i}-1)}\Big\}\Upsilon_{i}^{00}\Upsilon_{i}^{01}\\
&\quad \quad \quad \quad \quad -\frac{2n_{i}^{2}}{N^{2}m_{i}(n_{i}-m_{i}) }\Upsilon_{i}^{00}\Upsilon_{i}^{10}-\frac{2n_{i}^{2}}{N^{2}m_{i}(n_{i}-m_{i}) }\Upsilon_{i}^{00}\Upsilon_{i}^{11}\\
&\quad \quad \quad \quad \quad -\frac{2n_{i}^{2}}{N^{2}m_{i}(n_{i}-m_{i}) }\Upsilon_{i}^{01}\Upsilon_{i}^{10}-\frac{2n_{i}^{2}}{N^{2}m_{i}(n_{i}-m_{i}) }\Upsilon_{i}^{01}\Upsilon_{i}^{11}\\
& \quad \quad \quad \quad \quad + 2\Big\{ \frac{n_{i}^{2}}{N^{2}m_{i}^{2} }+\chi^{2}_{1, 1-\alpha} \cdot \frac{n_{i}^{2}}{N^{2}m_{i}^{2}(m_{i}-1)}\Big\}\Upsilon_{i}^{10}\Upsilon_{i}^{11}\\
&\quad \quad \quad \quad \quad -\frac{\chi^{2}_{1, 1-\alpha}\cdot n_{i}^{2} }{N^{2}(n_{i}-m_{i})(n_{i}-m_{i}-1)}(\Upsilon_{i}^{00}+\Upsilon_{i}^{01})- \frac{\chi^{2}_{1, 1-\alpha}\cdot n_{i}^{2} }{N^{2}m_{i}(m_{i}-1)}(\Upsilon_{i}^{10}+\Upsilon_{i}^{11})\Big] \\
& \quad \quad  + \sum_{i\neq i^{\prime} }\Big\{ \frac{n_{i}n_{i^{\prime}} }{N^{2}(n_{i}-m_{i})(n_{i^{\prime} }-m_{i^{\prime} }) } \Upsilon_{i}^{00}\Upsilon_{i^{\prime} }^{00}+\frac{n_{i}n_{i^{\prime}} }{N^{2}(n_{i}-m_{i})(n_{i^{\prime} }-m_{i^{\prime} }) } \Upsilon_{i}^{00}\Upsilon_{i^{\prime} }^{01}\\
&\quad \quad \quad \quad \quad \quad -\frac{n_{i}n_{i^{\prime}} }{N^{2}(n_{i}-m_{i})m_{i^{\prime} } } \Upsilon_{i}^{00}\Upsilon_{i^{\prime} }^{10}-\frac{n_{i}n_{i^{\prime}} }{N^{2}(n_{i}-m_{i})m_{i^{\prime} } } \Upsilon_{i}^{00}\Upsilon_{i^{\prime} }^{11}\\
&\quad \quad \quad \quad \quad \quad +\frac{n_{i}n_{i^{\prime}} }{N^{2}(n_{i}-m_{i})(n_{i^{\prime} }-m_{i^{\prime} }) } \Upsilon_{i}^{01}\Upsilon_{i^{\prime} }^{00}+\frac{n_{i}n_{i^{\prime}} }{N^{2}(n_{i}-m_{i})(n_{i^{\prime} }-m_{i^{\prime} }) } \Upsilon_{i}^{01}\Upsilon_{i^{\prime} }^{01}\\
& \quad \quad \quad \quad \quad \quad -\frac{n_{i}n_{i^{\prime}} }{N^{2}(n_{i}-m_{i})m_{i^{\prime} } } \Upsilon_{i}^{01}\Upsilon_{i^{\prime} }^{10}-\frac{n_{i}n_{i^{\prime}} }{N^{2}(n_{i}-m_{i})m_{i^{\prime} } } \Upsilon_{i}^{01}\Upsilon_{i^{\prime} }^{11}\\
&\quad \quad \quad \quad \quad \quad-\frac{n_{i }n_{i^{\prime} } }{N^{2}m_{i}(n_{i^{\prime} }-m_{i^{\prime} })} \Upsilon_{i}^{10}\Upsilon_{i^{\prime}}^{00}- \frac{n_{i }n_{i^{\prime} } }{N^{2}m_{i}(n_{i^{\prime} }-m_{i^{\prime} })} \Upsilon_{i}^{10}\Upsilon_{i^{\prime}}^{01}\\
&\quad \quad \quad \quad \quad \quad +\frac{n_{i }n_{i^{\prime} } }{N^{2}m_{i}m_{i^{\prime} }} \Upsilon_{i}^{10}\Upsilon_{i^{\prime}}^{10}+\frac{n_{i }n_{i^{\prime} } }{N^{2}m_{i}m_{i^{\prime} }} \Upsilon_{i}^{10}\Upsilon_{i^{\prime}}^{11}-\frac{n_{i }n_{i^{\prime} } }{N^{2}m_{i}(n_{i^{\prime} }-m_{i^{\prime} })} \Upsilon_{i}^{11}\Upsilon_{i^{\prime}}^{00}\\
&  \quad \quad \quad \quad \quad \quad - \frac{n_{i }n_{i^{\prime} } }{N^{2}m_{i}(n_{i^{\prime} }-m_{i^{\prime} })} \Upsilon_{i}^{11}\Upsilon_{i^{\prime}}^{01}+\frac{n_{i }n_{i^{\prime} } }{N^{2}m_{i}m_{i^{\prime} }} \Upsilon_{i}^{11}\Upsilon_{i^{\prime}}^{10}+\frac{n_{i }n_{i^{\prime} } }{N^{2}m_{i}m_{i^{\prime} }} \Upsilon_{i}^{11}\Upsilon_{i^{\prime}}^{11}\Big\},
\end{align*}
which enters into the quadratic constraint for the integer program (P3), of which a standard form is given in Web Appendix B.3. The above equation can also be written as
\begin{align*}
&\{T_{\text{Neyman}}(\mathbf{Z}, \mathbf{Y})\}^{2}-\chi^{2}_{1, 1-\alpha} \cdot \widehat{\text{Var}}\{T_{\text{Neyman}}(\mathbf{Z}, \mathbf{Y})\}\\
    &=\Big[ \sum_{s=1}^{S}\sum_{p=1}^{\widetilde{N}_{S} }d_{sp}\Big\{\frac{\widetilde{n}_{s}}{N\widetilde{m}_{s}}(\Delta_{sp}^{10}+\Delta_{sp}^{11}) -\frac{\widetilde{n}_{s}}{N(\widetilde{n}_{s}-\widetilde{m}_{s})}(\Delta_{sp}^{00}+\Delta_{sp}^{01}) \Big\} \Big]^{2}\\
&\quad \quad \quad -\chi^{2}_{1, 1-\alpha} \cdot \sum_{s=1}^{S}\sum_{p=1}^{\widetilde{N}_{S}} d_{sp}\Big(\frac{\widetilde{n}_{s}}{N}\Big)^{2} \Big\{ \frac{\Delta_{sp}^{10}+\Delta_{sp}^{11} }{\widetilde{m}_{s}(\widetilde{m}_{s}-1)}-\frac{(\Delta_{sp}^{10}+\Delta_{sp}^{11})^{2} }{\widetilde{m}_{s}^{2}(\widetilde{m}_{s}-1)}\\
&\quad \quad \quad \quad \quad \quad \quad \quad \quad \quad + \frac{\Delta_{sp}^{00}+\Delta_{sp}^{01} }{(\widetilde{n}_{s}-\widetilde{m}_{s})(\widetilde{n}_{s}-\widetilde{m}_{s}-1)}-\frac{(\Delta_{sp}^{00}+\Delta_{sp}^{01})^{2} }{(\widetilde{n}_{s}-\widetilde{m}_{s})^{2}(\widetilde{n}_{s}-\widetilde{m}_{s}-1)}\Big\}\\
&=\sum_{s=1}^{S}\sum_{p=1}^{\widetilde{N}_{s}}\sum_{s^{\prime}=1}^{S}\sum_{p^{\prime}=1}^{\widetilde{N}_{s^{\prime} }}d_{sp}d_{s^{\prime}p^{\prime}} \Big[ \Big\{\frac{\widetilde{n}_{s}}{N\widetilde{m}_{s}}(\Delta_{sp}^{10}+\Delta_{sp}^{11})-\frac{\widetilde{n}_{s}}{N(\widetilde{n}_{s}-\widetilde{m}_{s})}(\Delta_{sp}^{00}+\Delta_{sp}^{01}) \Big\}\\
          &\quad \quad \quad \quad \quad \quad \quad \quad \quad \quad \quad \quad \quad \quad \times \Big\{\frac{\widetilde{n}_{s^{\prime} }}{N\widetilde{m}_{s^{\prime} }}(\Delta_{s^{\prime}p^{\prime}}^{10}+\Delta_{s^{\prime}p^{\prime} }^{11}) -\frac{\widetilde{n}_{s^{\prime} }}{N(\widetilde{n}_{s^{\prime} }-\widetilde{m}_{s^{\prime} })}(\Delta_{s^{\prime}p^{\prime}}^{00}+\Delta_{s^{\prime}p^{\prime}}^{01}) \Big\}\Big]\\
   &\quad \quad \quad -\chi^{2}_{1, 1-\alpha} \cdot \sum_{s=1}^{S}\sum_{p=1}^{\widetilde{N}_{S}} d_{sp}\Big(\frac{\widetilde{n}_{s}}{N}\Big)^{2} \Big\{ \frac{\Delta_{sp}^{10}+\Delta_{sp}^{11} }{\widetilde{m}_{s}(\widetilde{m}_{s}-1)}-\frac{(\Delta_{sp}^{10}+\Delta_{sp}^{11})^{2} }{\widetilde{m}_{s}^{2}(\widetilde{m}_{s}-1)}\\
&\quad \quad \quad \quad \quad \quad \quad \quad \quad \quad + \frac{\Delta_{sp}^{00}+\Delta_{sp}^{01} }{(\widetilde{n}_{s}-\widetilde{m}_{s})(\widetilde{n}_{s}-\widetilde{m}_{s}-1)}-\frac{(\Delta_{sp}^{00}+\Delta_{sp}^{01})^{2} }{(\widetilde{n}_{s}-\widetilde{m}_{s})^{2}(\widetilde{n}_{s}-\widetilde{m}_{s}-1)}\Big\},
\end{align*}
which enters into the quadratic constraint for the integer program (P4), of which a standard form is given in Web Appendix B.4.

\newpage
 \section*{Web Appendix B: Detailed Formulations of the Related Integer Programs for Computing Warning Accuracy in Various Cases}
 
\subsection*{B.1: Warning accuracy with Fisher's sharp null (type I randomization designs)}

We write the following integer quadratically constrained linear program (IQCLP)
\begin{equation*}
     \begin{split}
        \underset{\mathbf{\Upsilon}\in  \mathbb{Z}^{4I}} {\text{max}} \quad &\frac{1}{N}\sum_{i=1}^{I}(\Upsilon_{i}^{01}+\Upsilon_{i}^{11}-\Upsilon_{i}^{00}-\Upsilon_{i}^{10})+\frac{1}{N}\sum_{i=1}^{I}\sum_{j=1}^{n_{i}}(1-Y_{ij}^{*}) \quad \quad (\text{P}1)\\
         \text{s.t.}\quad & \Big\{ \sum_{i=1}^{I}(\Upsilon_{i}^{10}+\Upsilon_{i}^{11})-\sum_{i=1}^{I}\frac{m_{i}}{n_{i}} \breve{\Upsilon}_{i} \Big\}^{2}- \chi^{2}_{1, 1-\alpha} \cdot \sum_{i=1}^{I}\frac{m_{i} \breve{\Upsilon}_{i}  (n_{i}-\breve{\Upsilon}_{i})(n_{i}-m_{i})}{n_{i}^{2}(n_{i}-1)}\leq 0, \\
        &0 \leq \Upsilon_{i}^{00} \leq \sum_{j=1}^{n_{i}}(1-Z_{ij})(1-Y_{ij}^{*}), \quad \forall i \\  
        &0 \leq \Upsilon_{i}^{01} \leq \sum_{j=1}^{n_{i}}(1-Z_{ij})Y_{ij}^{*}, \quad \forall i \\
        &0 \leq \Upsilon_{i}^{10} \leq \sum_{j=1}^{n_{i}}Z_{ij}(1-Y_{ij}^{*}), \quad \forall i \\
        &0 \leq \Upsilon_{i}^{11} \leq \sum_{j=1}^{n_{i}}Z_{ij}Y_{ij}^{*}, \quad \forall i \\ 
     \end{split}
 \end{equation*}
in a standard form
\begin{equation*}
     \begin{split}
        \underset{\mathbf{x}}{\text{max}} \quad & \mathbf{q}^T \mathbf{x}+c\\
         \text{s.t.}\quad & \mathbf{x}^T \mathbf{Q}_{1} \mathbf{x} + \mathbf{q}_{1}^T \mathbf{x} \leq 0, \\
        & \mathbf{l}\leq \mathbf{x} \leq \mathbf{u}, \\
        & \text{All elements of} ~\mathbf{x}~ \text{are integers.}
     \end{split}
 \end{equation*}
Specifically, we have
\begin{itemize}
    \item decision variables: $\mathbf{x}=\mathbf{\Upsilon}=(\Upsilon_{1}^{00}, \Upsilon_{1}^{01}, \Upsilon_{1}^{10}, \Upsilon_{1}^{11}, \dots, \Upsilon_{I}^{00}, \Upsilon_{I}^{01}, \Upsilon_{I}^{10}, \Upsilon_{I}^{11})$.
    \item objective function: $\mathbf{q}^{T}\mathbf{x}+c$ where
    \begin{equation*}
        \mathbf{q}=\Big( -\frac{1}{N}, \frac{1}{N}, -\frac{1}{N}, \frac{1}{N}, \dots,  -\frac{1}{N}, \frac{1}{N}, -\frac{1}{N}, \frac{1}{N} \Big) \text{ and } c=\frac{1}{N}\sum_{i=1}^{I}\sum_{j=1}^{n_{i}}(1-Y_{ij}^{*}). 
    \end{equation*}
    \item quadratic constraint: $\mathbf{x}^T \mathbf{Q}_{1} \mathbf{x} + \mathbf{q_{1}}^T \mathbf{x} \leq 0$ where $\mathbf{Q}_{1}=(Q_{1, st})_{4I \times 4I}$ is a $4I \times 4I$ matrix. Suppose that $s=4(i-1)+k$ and $t=4(i^{\prime}-1)+k^{\prime}$ for some integers $i, i^{\prime} \in \{1, \dots, I\}$ and $k, k^{\prime} \in  \{1, 2, 3, 4\}$. Then we have:
    \begin{enumerate}
        \item If $(s,t)$ satisfies one of the following conditions: 1) $i=i^{\prime}$ and $k=k^{\prime}=1$; 2) $i=i^{\prime}$ and $k=k^{\prime}=2$; 3) $i=i^{\prime}$ and $k=1, k^{\prime}=2$; 4) $i=i^{\prime}$ and $k=2, k^{\prime}=1$, we have 
        \begin{equation*}
            Q_{1,st}=\frac{m_{i}^{2}}{n_{i}^{2}}+\chi^{2}_{1, 1-\alpha}\cdot \frac{m_{i}(n_{i}-m_{i}) }{n_{i}^{2}(n_{i}-1)}.
        \end{equation*}
        \item If $(s,t)$ satisfies one of the following conditions: 1) $i=i^{\prime}$ and $k=k^{\prime}=3$; 2) $i=i^{\prime}$ and $k=k^{\prime}=4$; 3) $i=i^{\prime}$ and $k=3, k^{\prime}=4$; 4) $i=i^{\prime}$ and $k=4, k^{\prime}=3$, we have 
        \begin{equation*}
            Q_{1, st}=\Big(1-\frac{m_{i}}{n_{i}}\Big)^{2}+\chi^{2}_{1, 1-\alpha}\cdot \frac{m_{i}(n_{i}-m_{i})}{n_{i}^{2}(n_{i}-1)}.
        \end{equation*}
        \item If $(s,t)$ satisfies one of the following conditions: 1) $i=i^{\prime}$ and $k=1, k^{\prime}=3$; 2) $i=i^{\prime}$ and $k=3, k^{\prime}=1$; 3) $i=i^{\prime}$ and $k=1, k^{\prime}=4$; 4) $i=i^{\prime}$ and $k=4, k^{\prime}=1$; 5) $i=i^{\prime}$ and $k=2, k^{\prime}=3$; 6) $i=i^{\prime}$ and $k=3, k^{\prime}=2$; 7) $i=i^{\prime}$ and $k=2, k^{\prime}=4$; 8) $i=i^{\prime}$ and $k=4, k^{\prime}=2$; we have 
        \begin{equation*}
            Q_{1,st}=- \frac{m_{i}}{n_{i}}\Big(1-\frac{m_{i}}{n_{i}}\Big)+\chi^{2}_{1, 1-\alpha}\cdot \frac{m_{i}(n_{i}-m_{i})}{n_{i}^{2}(n_{i}-1)}.
        \end{equation*}
        \item If $(s,t)$ satisfies one of the following conditions: 1) $i\neq i^{\prime}$ and $k=k^{\prime}=1$; 2) $i\neq i^{\prime}$ and $k=1, k^{\prime}=2$; 3) $i\neq i^{\prime}$ and $k=2, k^{\prime}=1$; 4) $i\neq i^{\prime}$ and $k=k^{\prime}=2$, we have
        \begin{equation*}
            Q_{1, st}=\frac{m_{i}}{n_{i}}\frac{m_{i^{\prime} }}{n_{i^{\prime}}}.
        \end{equation*}
        \item If $(s,t)$ satisfies one of the following conditions: 1) $i\neq i^{\prime}$ and $k=1, k^{\prime}=3$; 2) $i\neq i^{\prime}$ and $k=1, k^{\prime}=4$; 3) $i\neq i^{\prime}$ and $k=2, k^{\prime}=3$; 4) $i\neq i^{\prime}$ and $k=2, k^{\prime}=4$, we have
        \begin{equation*}
            Q_{1, st}=-\frac{m_{i}}{n_{i}}\Big(1-\frac{m_{i^{\prime} }}{n_{i^{\prime}}}\Big).
        \end{equation*}
        \item If $(s,t)$ satisfies one of the following conditions: 1) $i\neq i^{\prime}$ and $k=3, k^{\prime}=1$; 2) $i\neq i^{\prime}$ and $k=3, k^{\prime}=2$; 3) $i\neq i^{\prime}$ and $k=4, k^{\prime}=1$; 4) $i\neq i^{\prime}$ and $k=4, k^{\prime}=2$, we have
        \begin{equation*}
            Q_{1, st}=-\Big(1-\frac{m_{i}}{n_{i}}\Big)\frac{m_{i^{\prime} }}{n_{i^{\prime}}}.
        \end{equation*}
         \item If $(s,t)$ satisfies one of the following conditions: 1) $i\neq i^{\prime}$ and $k=3, k^{\prime}=3$; 2) $i\neq i^{\prime}$ and $k=3, k^{\prime}=4$; 3) $i\neq i^{\prime}$ and $k=4, k^{\prime}=3$; 4) $i\neq i^{\prime}$ and $k=4, k^{\prime}=4$, we have
        \begin{equation*}
            Q_{1, st}=\Big(1-\frac{m_{i}}{n_{i}}\Big)\Big(1-\frac{m_{i^{\prime} }}{n_{i^{\prime}}}\Big).
        \end{equation*}
    \end{enumerate}
            We have $\mathbf{q}_{1}=(q_{1, 1}, \dots, q_{1, 4I})$ is a $4I$-dimensional vector with
        \begin{equation*}
            q_{1, s}=-\chi^{2}_{1, 1-\alpha} \cdot \frac{m_{i}n_{i}(n_{i}-m_{i})}{n_{i}^{2}(n_{i}-1)}, \ \text{for $s=4(i-1)+k, i \in \{1,\dots, I\}, k\in \{1,2,3,4\}$.} 
        \end{equation*}
        
        \item box constraints: $\mathbf{l}\leq \mathbf{\Upsilon}\leq \mathbf{u}$, where $\mathbf{l}=\mathbf{0}$ and 
        \begin{equation*}
            \mathbf{u}=(M_{1}^{00}, M_{1}^{01}, M_{1}^{10}, M_{1}^{11}, \dots, M_{I}^{00}, M_{I}^{01}, M_{I}^{10}, M_{I}^{11})
        \end{equation*} 
        is a $4I$-dimensional vector with $M_{i}^{00}=\sum_{j=1}^{n_{i}}(1-Z_{ij})(1-Y_{ij}^{*}), M_{i}^{01}=\sum_{j=1}^{n_{i}}(1-Z_{ij})Y_{ij}^{*}, M_{i}^{10}=\sum_{j=1}^{n_{i}}Z_{ij}(1-Y_{ij}^{*})$, and $M_{i}^{11}=\sum_{j=1}^{n_{i}}Z_{ij}Y_{ij}^{*}$.
        \item integer constraints: all $4I$ elements of $\mathbf{x}$ are integers. 
\end{itemize}

\subsection*{B.2: Warning accuracy with Fisher's sharp null (type II randomization designs)}

We write the following integer quadratically constrained linear program
\begin{equation*}
     \begin{split}
        \underset{d_{sp}\in \mathbb{Z}}{\text{max}} \quad & \frac{1}{N}\sum_{s=1}^{S}\sum_{p=1}^{\widetilde{N}_{s}}d_{sp}(\Delta_{sp}^{01}+\Delta_{sp}^{11}-\Delta_{sp}^{00}-\Delta_{sp}^{10})+\frac{1}{N}\sum_{i=1}^{I}\sum_{j=1}^{n_{i}}(1-Y_{ij}^{*}) \quad \quad (\text{P}2)\\
         \text{s.t. }\quad  & \Big\{ \sum_{s=1}^{S}\sum_{p=1}^{\widetilde{N}_{s}}d_{sp}(\Delta_{sp}^{10}+\Delta_{sp}^{11})-\sum_{s=1}^{S}\sum_{p=1}^{\widetilde{N}_{s}}d_{sp}\Big (\frac{\widetilde{m}_{s}}{\widetilde{n}_{s}}\cdot \breve{\Delta}_{sp}\Big) \Big\}^{2}\\
   &\quad \quad \quad \quad \quad \quad  - \chi^{2}_{1, 1-\alpha} \cdot \sum_{s=1}^{S}\sum_{p=1}^{\widetilde{N}_{s}}d_{sp}\frac{\widetilde{m}_{s} \breve{\Delta}_{sp} (\widetilde{n}_{s}-\breve{\Delta}_{sp}) (\widetilde{n}_{s}-\widetilde{m}_{s})}{\widetilde{n}_{s}^{2}(\widetilde{n}_{s}-1)}\leq 0, \\
        & \sum_{p=1}^{\widetilde{N}_{s}} d_{sp} = P_{s}, \quad \forall s \\
        & d_{sp}\geq 0, \quad \forall s, p
     \end{split}
 \end{equation*}
in a standard form
\begin{equation*}
     \begin{split}
        \underset{\mathbf{x}}{\text{max}} \quad & \mathbf{q}^T \mathbf{x}+c\\
         \text{s.t.}\quad & \mathbf{x}^T \mathbf{Q}_{1} \mathbf{x} + \mathbf{q}_{1}^T \mathbf{x} \leq 0, \\
          & \mathbf{q}_{2s}^{T}\mathbf{x} = P_{s}, \quad \forall s \\
        &  \mathbf{x} \geq \mathbf{0}, \\
        & \text{All elements of} ~\mathbf{x}~ \text{are integers.}
     \end{split}
 \end{equation*}
We have: 
\begin{itemize}
    \item decision variables: $\mathbf{x}=\mathbf{d}=(d_{11}, \dots, d_{S \widetilde{N}_{S}})$;
    \item objective function: $\mathbf{q}^{T}\mathbf{x}+c$ where
    \begin{equation*}
        \mathbf{q}=\Big(\frac{\Delta_{11}^{01}+\Delta_{11}^{11}-\Delta_{11}^{00}-\Delta_{11}^{10}}{N}, \dots, \frac{\Delta_{S\widetilde{N}_{S} }^{01}+\Delta_{S\widetilde{N}_{S}}^{11}-\Delta_{S\widetilde{N}_{S}}^{00}-\Delta_{S\widetilde{N}_{S} }^{10}}{N} \Big)
    \end{equation*}
    and
    \begin{equation*}
        c=\frac{1}{N}\sum_{i=1}^{I}\sum_{j=1}^{n_{i}}(1-Y_{ij}^{*}).
    \end{equation*}
\item quadratic constraint: $\mathbf{x}^T \mathbf{Q}_{1} \mathbf{x} + \mathbf{q_{1}}^T \mathbf{x} \leq 0$ where $\mathbf{Q}_{1}=(Q_{1, rt})_{\widetilde{N} \times \widetilde{N}}$ is a $\widetilde{N} \times \widetilde{N}$ matrix ($\widetilde{N}=\sum_{s=1}^{S}\widetilde{N}_{s}$). Suppose that $r$ corresponds to the $p$-th unique $2 \times 2 \times 2$ table for the $s$-th unique $2 \times 2$ table $\Lambda_{[s]}$, and $t$ corresponds to the $p^{\prime}$-th unique $2 \times 2 \times 2$ table for the $s^{\prime}$-th unique $2 \times 2$ table $\Lambda_{[s^{\prime}]}$. Then we have:
\begin{equation*}
Q_{1,rt}=\big \{ \Delta_{sp}^{10}+\Delta_{sp}^{11}-\frac{\widetilde{m}_{s}}{\widetilde{n}_{s}}(\Delta_{sp}^{00}+\Delta_{sp}^{01}+\Delta_{sp}^{10}+\Delta_{sp}^{11})\big \}\big \{ \Delta_{s^{\prime}p^{\prime}}^{10}+\Delta_{s^{\prime}p^{\prime}}^{11}-\frac{\widetilde{m}_{s^{\prime} }}{\widetilde{n}_{s^{\prime} }}(\Delta_{s^{\prime}p^{\prime}}^{00}+\Delta_{s^{\prime}p^{\prime}}^{01}+\Delta_{s^{\prime}p^{\prime}}^{10}+\Delta_{s^{\prime}p^{\prime}}^{11})\big \}.
\end{equation*}
    We have $\mathbf{q}_{1}=(q_{1,11}, \dots, q_{1,S\widetilde{N}_{S}})$ is a $\widetilde{N}$-dimensional vector where
        \begin{equation*}
            q_{1,sp}= - \chi^{2}_{1, 1-\alpha} \cdot \frac{\widetilde{m}_{s}(\Delta_{sp}^{00}+\Delta_{sp}^{01}+\Delta_{sp}^{10}+\Delta_{sp}^{11})(\widetilde{n}_{s}-\Delta_{sp}^{00}-\Delta_{sp}^{01}-\Delta_{sp}^{10}-\Delta_{sp}^{11})(\widetilde{n}_{s}-\widetilde{m}_{s})}{\widetilde{n}_{s}^{2}(\widetilde{n}_{s}-1)}.
        \end{equation*}
                \item linear constraints: $\mathbf{q}_{2s}^{T} \mathbf{x}=P_{s}$, where $\mathbf{q}_{2s}$ is the zero-one indicator vector for the $\widetilde{N}_{s}$ unique $2 \times 2 \times 2$ tables of $\Lambda_{[s]}$.
        \item box constraints: $d_{sp}\geq 0$ for all $s$ and $p$.                \item integer constraints: all $\widetilde{N}$ elements of $\mathbf{x}$ are integers. 
\end{itemize}

\subsection*{B.3: Warning accuracy with Neyman's weak null (type I randomization designs)}

Following a similar argument as in Section 4.2.1 of the main text, we can reformulate the integer program ($\text{P}0^{\prime}$) as the following integer quadratically constrained linear program
\begin{equation*}
     \begin{split}
        \underset{\mathbf{\Upsilon}\in \mathbb{Z}^{4I}} {\text{max}} \quad &\frac{1}{N}\sum_{i=1}^{I}(\Upsilon_{i}^{01}+\Upsilon_{i}^{11}-\Upsilon_{i}^{00}-\Upsilon_{i}^{10})+\frac{1}{N}\sum_{i=1}^{I}\sum_{j=1}^{n_{i}}(1-Y_{ij}^{*}) \quad \quad (\text{P}3)\\
         \text{s.t.}\quad &\Big[ \sum_{i=1}^{I}\Big\{\frac{n_{i}}{Nm_{i}}(\Upsilon_{i}^{10}+\Upsilon_{i}^{11}) -\frac{n_{i}}{N(n_{i}-m_{i})}(\Upsilon_{i}^{00}+\Upsilon_{i}^{01}) \Big\} \Big]^{2}\\
&\quad \quad \quad \quad \quad -\chi^{2}_{1, 1-\alpha} \cdot \sum_{i=1}^{I} \Big(\frac{n_{i}}{N}\Big)^{2} \Big\{ \frac{\Upsilon_{i}^{10}+\Upsilon_{i}^{11} }{m_{i}(m_{i}-1)}-\frac{(\Upsilon_{i}^{10}+\Upsilon_{i}^{11})^{2} }{m_{i}^{2}(m_{i}-1)}\\
&\quad \quad \quad \quad \quad \quad \quad + \frac{\Upsilon_{i}^{00}+\Upsilon_{i}^{01} }{(n_{i}-m_{i})(n_{i}-m_{i}-1)}-\frac{(\Upsilon_{i}^{00}+\Upsilon_{i}^{01})^{2} }{(n_{i}-m_{i})^{2}(n_{i}-m_{i}-1)}\Big\}\leq 0, \\
              &0 \leq \Upsilon_{i}^{00} \leq \sum_{j=1}^{n_{i}}(1-Z_{ij})(1-Y_{ij}^{*}), \quad \forall i \\  
        &0 \leq \Upsilon_{i}^{01} \leq \sum_{j=1}^{n_{i}}(1-Z_{ij})Y_{ij}^{*}, \quad \forall i \\
        &0 \leq \Upsilon_{i}^{10} \leq \sum_{j=1}^{n_{i}}Z_{ij}(1-Y_{ij}^{*}), \quad \forall i \\
        &0 \leq \Upsilon_{i}^{11} \leq \sum_{j=1}^{n_{i}}Z_{ij}Y_{ij}^{*}, \quad \forall i 
     \end{split}
 \end{equation*}
 which can be written as the following standard form 
 \begin{equation*}
     \begin{split}
        \underset{\mathbf{x}}{\text{max}} \quad & \mathbf{q}^T \mathbf{x}+c\\
         \text{s.t.}\quad & \mathbf{x}^T \mathbf{Q}_{1} \mathbf{x}+\mathbf{q}_{1}^{T}\mathbf{x} \leq 0, \\
         & \mathbf{l}\leq \mathbf{x}\leq \mathbf{u}, \\
        & \text{All elements of} ~\mathbf{x}~ \text{are integers.}
     \end{split}
 \end{equation*}
Specifically, we have: 
\begin{itemize}
    \item decision variables: $\mathbf{x}=\mathbf{\Upsilon}=(\Upsilon_{1}^{00}, \Upsilon_{1}^{01}, \Upsilon_{1}^{10}, \Upsilon_{1}^{11}, \dots, \Upsilon_{I}^{00}, \Upsilon_{I}^{01}, \Upsilon_{I}^{10}, \Upsilon_{I}^{11})$;
    \item objective function: $\mathbf{q}^{T}\mathbf{x}+c$ where
    \begin{equation*}
        \mathbf{q}=\Big( -\frac{1}{N}, \frac{1}{N}, -\frac{1}{N}, \frac{1}{N}, \dots,  -\frac{1}{N}, \frac{1}{N}, -\frac{1}{N}, \frac{1}{N} \Big) \text{ and } c=\frac{1}{N}\sum_{i=1}^{I}\sum_{j=1}^{n_{i}}(1-Y_{ij}^{*}). 
    \end{equation*}
\item quadratic constraint: $\mathbf{x}^T \mathbf{Q}_{1} \mathbf{x} + \mathbf{q_{1}}^T \mathbf{x} \leq 0$ where $\mathbf{Q}_{1}=(Q_{1, st})_{4I \times 4I}$ is a $4I \times 4I$ matrix. Suppose that $s=4(i-1)+k$ and $t=4(i^{\prime}-1)+k^{\prime}$ for some integers $i, i^{\prime} \in \{1, \dots, I\}$ and $k, k^{\prime} \in  \{1, 2, 3, 4\}$. Then we have:
    \begin{enumerate}
        \item If $(s,t)$ satisfies one of the following conditions: 1) $i=i^{\prime}$ and $k=k^{\prime}=1$; 2) $i=i^{\prime}$ and $k=k^{\prime}=2$; 3) $i=i^{\prime}$ and $k=1, k^{\prime}=2$; 4) $i=i^{\prime}$ and $k=2, k^{\prime}=1$, we have 
        \begin{equation*}
            Q_{1,st}=\frac{n_{i}^{2}}{N^{2}(n_{i}-m_{i})^{2} }+\chi^{2}_{1, 1-\alpha} \cdot \frac{n_{i}^{2}}{N^{2}(n_{i}-m_{i})^{2}(n_{i}-m_{i}-1)}.
        \end{equation*}
        \item If $(s,t)$ satisfies one of the following conditions: 1) $i=i^{\prime}$ and $k=k^{\prime}=3$; 2) $i=i^{\prime}$ and $k=k^{\prime}=4$; 3) $i=i^{\prime}$ and $k=3, k^{\prime}=4$; 4) $i=i^{\prime}$ and $k=4, k^{\prime}=3$, we have 
        \begin{equation*}
            Q_{1, st}=\frac{n_{i}^{2}}{N^{2}m_{i}^{2} }+\chi^{2}_{1, 1-\alpha} \cdot \frac{n_{i}^{2}}{N^{2}m_{i}^{2}(m_{i}-1)}.
        \end{equation*}
        \item If $(s,t)$ satisfies one of the following conditions: 1) $i=i^{\prime}$ and $k=1, k^{\prime}=3$; 2) $i=i^{\prime}$ and $k=3, k^{\prime}=1$; 3) $i=i^{\prime}$ and $k=1, k^{\prime}=4$; 4) $i=i^{\prime}$ and $k=4, k^{\prime}=1$; 5) $i=i^{\prime}$ and $k=2, k^{\prime}=3$; 6) $i=i^{\prime}$ and $k=3, k^{\prime}=2$; 7) $i=i^{\prime}$ and $k=2, k^{\prime}=4$; 8) $i=i^{\prime}$ and $k=4, k^{\prime}=2$; we have 
        \begin{equation*}
            Q_{1,st}=-\frac{n_{i}^{2}}{N^{2}m_{i}(n_{i}-m_{i}) }.
        \end{equation*}
        \item If $(s,t)$ satisfies one of the following conditions: 1) $i\neq i^{\prime}$ and $k=k^{\prime}=1$; 2) $i\neq i^{\prime}$ and $k=1, k^{\prime}=2$; 3) $i\neq i^{\prime}$ and $k=2, k^{\prime}=1$; 4) $i\neq i^{\prime}$ and $k=k^{\prime}=2$, we have
        \begin{equation*}
            Q_{1, st}= \frac{n_{i}n_{i^{\prime}} }{N^{2}(n_{i}-m_{i})(n_{i^{\prime} }-m_{i^{\prime} })}.
        \end{equation*}
        \item If $(s,t)$ satisfies one of the following conditions: 1) $i\neq i^{\prime}$ and $k=1, k^{\prime}=3$; 2) $i\neq i^{\prime}$ and $k=1, k^{\prime}=4$; 3) $i\neq i^{\prime}$ and $k=2, k^{\prime}=3$; 4) $i\neq i^{\prime}$ and $k=2, k^{\prime}=4$, we have
        \begin{equation*}
            Q_{1, st}=-\frac{n_{i}n_{i^{\prime}} }{N^{2}(n_{i}-m_{i})m_{i^{\prime} } }.
        \end{equation*}
        \item If $(s,t)$ satisfies one of the following conditions: 1) $i\neq i^{\prime}$ and $k=3, k^{\prime}=1$; 2) $i\neq i^{\prime}$ and $k=3, k^{\prime}=2$; 3) $i\neq i^{\prime}$ and $k=4, k^{\prime}=1$; 4) $i\neq i^{\prime}$ and $k=4, k^{\prime}=2$, we have
        \begin{equation*}
            Q_{1, st}=- \frac{n_{i}n_{i^{\prime}} }{N^{2}m_{i}(n_{i^{\prime} }-m_{i^{\prime}})}.
        \end{equation*}
         \item If $(s,t)$ satisfies one of the following conditions: 1) $i\neq i^{\prime}$ and $k=3, k^{\prime}=3$; 2) $i\neq i^{\prime}$ and $k=3, k^{\prime}=4$; 3) $i\neq i^{\prime}$ and $k=4, k^{\prime}=3$; 4) $i\neq i^{\prime}$ and $k=4, k^{\prime}=4$, we have
        \begin{equation*}
            Q_{1, st}=\frac{n_{i }n_{i^{\prime} } }{N^{2}m_{i}m_{i^{\prime} }}.
        \end{equation*}
    \end{enumerate}
            We have $\mathbf{q}_{1}=(q_{1, 1}, \dots, q_{1, 4I})$ is a $4I$-dimensional vector where
        \begin{equation*}
            \text{$q_{1, s}=-\frac{\chi^{2}_{1, 1-\alpha}\cdot n_{i}^{2} }{N^{2}(n_{i}-m_{i})(n_{i}-m_{i}-1)}$ for $k=1,2$, and $q_{1, s}=-\frac{\chi^{2}_{1, 1-\alpha}\cdot n_{i}^{2} }{N^{2}m_{i}(m_{i}-1)}$ for $k=3,4$.}
        \end{equation*}
        Here $s=4(i-1)+k$ for $i\in \{1,\dots, I\}$ and $k\in \{1,2,3,4\}$.
        \item box constraints: $\mathbf{l}\leq \mathbf{\Upsilon}\leq \mathbf{u}$, where $\mathbf{l}=\mathbf{0}$ and 
        \begin{equation*}
        \mathbf{u}=(M_{1}^{00}, M_{1}^{01}, M_{1}^{10}, M_{1}^{11}, \dots, M_{I}^{00}, M_{I}^{01}, M_{I}^{10}, M_{I}^{11}) 
        \end{equation*}
        is a $4I$-dimensional vector with $M_{i}^{00}=\sum_{j=1}^{n_{i}}(1-Z_{ij})(1-Y_{ij}^{*}), M_{i}^{01}=\sum_{j=1}^{n_{i}}(1-Z_{ij})Y_{ij}^{*}, M_{i}^{10}=\sum_{j=1}^{n_{i}}Z_{ij}(1-Y_{ij}^{*})$, and $M_{i}^{11}=\sum_{j=1}^{n_{i}}Z_{ij}Y_{ij}^{*}$.
                \item integer constraints: all $4I$ elements of $\mathbf{x}$ are integers. 
\end{itemize}

\subsection*{B.4: Warning accuracy with Neyman's weak null (type II randomization designs)}

Following a similar argument as in Section 4.2.2 of the main text, we can reformulate the integer program ($\text{P}0^{\prime}$) as the following integer quadratically constrained linear program
\begin{equation*}
     \begin{split}
        \underset{d_{sp}\in \mathbb{Z} }{\text{max}} \quad & \frac{1}{N}\sum_{s=1}^{S}\sum_{p=1}^{\widetilde{N}_{s}}d_{sp}(\Delta_{sp}^{01}+\Delta_{sp}^{11}-\Delta_{sp}^{00}-\Delta_{sp}^{10})+\frac{1}{N}\sum_{i=1}^{I}\sum_{j=1}^{n_{i}}(1-Y_{ij}^{*}) \quad \quad (\text{P}4)\\
         \text{s.t. }\quad &\Big[ \sum_{s=1}^{S}\sum_{p=1}^{\widetilde{N}_{s} }d_{sp}\Big\{\frac{\widetilde{n}_{s}}{N\widetilde{m}_{s}}(\Delta_{sp}^{10}+\Delta_{sp}^{11}) -\frac{\widetilde{n}_{s}}{N(\widetilde{n}_{s}-\widetilde{m}_{s})}(\Delta_{sp}^{00}+\Delta_{sp}^{01}) \Big\} \Big]^{2}\\
&\quad  -\chi^{2}_{1, 1-\alpha} \cdot \sum_{s=1}^{S}\sum_{p=1}^{\widetilde{N}_{S}} d_{sp}\Big(\frac{\widetilde{n}_{s}}{N}\Big)^{2} \Big\{ \frac{\Delta_{sp}^{10}+\Delta_{sp}^{11} }{\widetilde{m}_{s}(\widetilde{m}_{s}-1)}-\frac{(\Delta_{sp}^{10}+\Delta_{sp}^{11})^{2} }{\widetilde{m}_{s}^{2}(\widetilde{m}_{s}-1)} \\
&\quad \quad \quad \quad \quad \quad \quad + \frac{\Delta_{sp}^{00}+\Delta_{sp}^{01} }{(\widetilde{n}_{s}-\widetilde{m}_{s})(\widetilde{n}_{s}-\widetilde{m}_{s}-1)}-\frac{(\Delta_{sp}^{00}+\Delta_{sp}^{01})^{2} }{(\widetilde{n}_{s}-\widetilde{m}_{s})^{2}(\widetilde{n}_{s}-\widetilde{m}_{s}-1)}\Big\}\leq 0, \\
        & \sum_{p=1}^{\widetilde{N}_{s}} d_{sp} = P_{s}, \quad \forall s \\
        & d_{sp}\geq 0, \quad \forall s, p 
     \end{split}
 \end{equation*}
which can be written as the following standard form
\begin{equation*}
     \begin{split}
        \underset{\mathbf{x}}{\text{max}} \quad & \mathbf{q}^T \mathbf{x}+c\\
         \text{s.t.}\quad & \mathbf{x}^T \mathbf{Q}_{1} \mathbf{x} + \mathbf{q}_{1}^T \mathbf{x} \leq 0, \\
          & \mathbf{q}_{2s}^{T}\mathbf{x} = P_{s}, \quad \forall s \\
        &  \mathbf{x} \geq \mathbf{0}, \\
        & \text{All elements of} ~\mathbf{x}~ \text{are integers.}
     \end{split}
 \end{equation*}
Specifically, we have: 
\begin{itemize}
    \item decision variables: $\mathbf{x}=\mathbf{d}=(d_{11}, \dots, d_{S \widetilde{N}_{S}})$;
    \item objective function: $\mathbf{q}^{T}\mathbf{x}+c$ where
    \begin{equation*}
        \mathbf{q}=\Big(\frac{\Delta_{11}^{01}+\Delta_{11}^{11}-\Delta_{11}^{00}-\Delta_{11}^{10}}{N}, \dots, \frac{\Delta_{S\widetilde{N}_{S} }^{01}+\Delta_{S\widetilde{N}_{S} }^{11}-\Delta_{S\widetilde{N}_{S} }^{00}-\Delta_{S\widetilde{N}_{S} }^{10}}{N} \Big)
    \end{equation*}
    and
    \begin{equation*}
        c=\frac{1}{N}\sum_{i=1}^{I}\sum_{j=1}^{n_{i}}(1-Y_{ij}^{*}).
    \end{equation*}
\item quadratic constraint: $\mathbf{x}^T \mathbf{Q}_{1} \mathbf{x} + \mathbf{q_{1}}^T \mathbf{x} \leq 0$ where $\mathbf{Q}_{1}=(Q_{1, rt})_{\widetilde{N} \times \widetilde{N}}$ is a $\widetilde{N} \times \widetilde{N}$ matrix ($\widetilde{N}=\sum_{s=1}^{S}\widetilde{N}_{s}$). Suppose that $r$ corresponds to the $p$-th unique $2 \times 2 \times 2$ table for the $s$-th unique $2 \times 2$ table $\Lambda_{[s]}$, and $t$ corresponds to the $p^{\prime}$-th unique $2 \times 2 \times 2$ table for the $s^{\prime}$-th unique $2 \times 2$ table $\Lambda_{[s^{\prime}]}$. Then we have
\begin{align*}
          Q_{1,rt}&=\Big\{\frac{\widetilde{n}_{s}}{N\widetilde{m}_{s}}(\Delta_{sp}^{10}+\Delta_{sp}^{11}) -\frac{\widetilde{n}_{s}}{N(\widetilde{n}_{s}-\widetilde{m}_{s})}(\Delta_{sp}^{00}+\Delta_{sp}^{01}) \Big\}\\
          &\quad \quad \quad \quad \quad \quad \times \Big\{\frac{\widetilde{n}_{s^{\prime} }}{N\widetilde{m}_{s^{\prime} }}(\Delta_{s^{\prime}p^{\prime}}^{10}+\Delta_{s^{\prime}p^{\prime} }^{11}) -\frac{\widetilde{n}_{s^{\prime} }}{N(\widetilde{n}_{s^{\prime} }-\widetilde{m}_{s^{\prime} })}(\Delta_{s^{\prime}p^{\prime}}^{00}+\Delta_{s^{\prime}p^{\prime}}^{01}) \Big\},
        \end{align*}
            and $\mathbf{q}_{1}=(q_{1,11}, \dots, q_{1,S\widetilde{N}_{S}})$ is a $\widetilde{N}$-dimensional vector where
        \begin{align*}
            q_{1,sp}&= -\chi^{2}_{1, 1-\alpha} \cdot \Big(\frac{\widetilde{n}_{s}}{N}\Big)^{2} \Big\{ \frac{\Delta_{sp}^{10}+\Delta_{sp}^{11} }{\widetilde{m}_{s}(\widetilde{m}_{s}-1)}-\frac{(\Delta_{sp}^{10}+\Delta_{sp}^{11})^{2} }{\widetilde{m}_{s}^{2}(\widetilde{m}_{s}-1)}\\
            &\quad \quad \quad \quad \quad \quad \quad \quad +\frac{\Delta_{sp}^{00}+\Delta_{sp}^{01} }{(\widetilde{n}_{s}-\widetilde{m}_{s})(\widetilde{n}_{s}-\widetilde{m}_{s}-1)}-\frac{(\Delta_{sp}^{00}+\Delta_{sp}^{01})^{2} }{(\widetilde{n}_{s}-\widetilde{m}_{s})^{2}(\widetilde{n}_{s}-\widetilde{m}_{s}-1)}\Big\}.
        \end{align*}
             \item linear constraints: $\mathbf{q}_{2s}^{T} \mathbf{x}=P_{s}$, where $\mathbf{q}_{2s}$ is the zero-one indicator vector for the $\widetilde{N}_{s}$ unique $2 \times 2 \times 2$ tables of $\Lambda_{[s]}$.
        \item box constraints: $d_{sp}\geq 0$ for all $s$ and $p$.                \item integer constraints: all $\widetilde{N}$ elements of $\mathbf{x}$ are integers. 
\end{itemize}

 \section*{Web Appendix C: More Details About Computing Sensitivity Weights and Sensitive Sets}
 
We here only consider Fisher's sharp null hypothesis (i.e., integer programs (P1) and (P2)) as the same method can also be applied for Neyman's weak null hypothesis. For type I randomization designs, let $\widetilde{\mathbf{\Upsilon}}=(\widetilde{\Upsilon}_{1}^{00}, \widetilde{\Upsilon}_{1}^{01}, \widetilde{\Upsilon}_{1}^{10}, \widetilde{\Upsilon}_{1}^{11}, \dots, \widetilde{\Upsilon}_{I}^{00}, \widetilde{\Upsilon}_{I}^{01}, \widetilde{\Upsilon}_{I}^{10}, \widetilde{\Upsilon}_{I}^{11})$ be an optimal solution to (P1). According to Definition 3 in the main text, we have 
\begin{align*}
    &W_{T}^{FP}=\frac{|\{ij: Z_{ij}=1, Y_{ij}^{*}=1, \widetilde{Y}_{ij}=0\}|}{|\{ij: Y_{ij}^{*}\neq \widetilde{Y}_{ij} \}|}=\frac{\sum_{i=1}^{I} (\Lambda_{i}^{11}-\widetilde{\Upsilon}_{i}^{11})}{\sum_{i=1}^{I} (\widetilde{\Upsilon}_{i}^{00}+\Lambda_{i}^{01}-\widetilde{\Upsilon}_{i}^{01}+\widetilde{\Upsilon}_{i}^{10}+\Lambda_{i}^{11}-\widetilde{\Upsilon}_{i}^{11})}, \\
    &W_{T}^{FN}=\frac{|\{ij: Z_{ij}=1, Y_{ij}^{*}=0, \widetilde{Y}_{ij}=1\}|}{|\{ij: Y_{ij}^{*}\neq \widetilde{Y}_{ij} \}|}=\frac{\sum_{i=1}^{I} \widetilde{\Upsilon}_{i}^{10}}{\sum_{i=1}^{I} (\widetilde{\Upsilon}_{i}^{00}+\Lambda_{i}^{01}-\widetilde{\Upsilon}_{i}^{01}+\widetilde{\Upsilon}_{i}^{10}+\Lambda_{i}^{11}-\widetilde{\Upsilon}_{i}^{11})},\\
    &W_{C}^{FP}=\frac{|\{ij: Z_{ij}=0, Y_{ij}^{*}=1, \widetilde{Y}_{ij}=0\}|}{|\{ij: Y_{ij}^{*}\neq \widetilde{Y}_{ij} \}|}=\frac{\sum_{i=1}^{I}(\Lambda_{i}^{01}-\widetilde{\Upsilon}_{i}^{01})}{\sum_{i=1}^{I} (\widetilde{\Upsilon}_{i}^{00}+\Lambda_{i}^{01}-\widetilde{\Upsilon}_{i}^{01}+\widetilde{\Upsilon}_{i}^{10}+\Lambda_{i}^{11}-\widetilde{\Upsilon}_{i}^{11})},\\
    &W_{C}^{FN}=\frac{|\{ij: Z_{ij}=0, Y_{ij}^{*}=0, \widetilde{Y}_{ij}=1\}|}{|\{ij: Y_{ij}^{*}\neq \widetilde{Y}_{ij} \}|}=\frac{\sum_{i=1}^{I} \widetilde{\Upsilon}_{i}^{00}}{\sum_{i=1}^{I} (\widetilde{\Upsilon}_{i}^{00}+\Lambda_{i}^{01}-\widetilde{\Upsilon}_{i}^{01}+\widetilde{\Upsilon}_{i}^{10}+\Lambda_{i}^{11}-\widetilde{\Upsilon}_{i}^{11})}.
\end{align*}
Note that after reformulating (P0) as (P1) or (P2), it is more natural to directly calculate a union of various sensitive sets. Specifically, let $\mathcal{S}$ be a sensitive set given from (P0), $G$ the permutation group over $\mathcal{I}$ defined in Section 4.1 of the main text and $g\mathcal{S}=\{g(ij): \widetilde{Y}_{ij}\neq Y_{ij}^{*}\}$ for $g\in G$, then we have 
  \begin{equation*}
      \bigcup_{g \in G}g\mathcal{S}=\bigcup_{i=1}^{I}\big\{ A_{i}^{00}\cup A_{i}^{01} \cup A_{i}^{10} \cup A_{i}^{11}\big\},
  \end{equation*}
  where $A_{i}^{00}=\{ij: Z_{ij}=0, Y_{ij}^{*}=0, \widetilde{\Upsilon}_{i}^{00}\neq 0, j=1,\dots, n_{i} \}$, $A_{i}^{01}=\{ij: Z_{ij}=0, Y_{ij}^{*}=1, \widetilde{\Upsilon}_{i}^{01}\neq \Lambda_{i}^{01}, j=1,\dots, n_{i} \}$, $A_{i}^{10}=\{ij: Z_{ij}=1, Y_{ij}^{*}=0, \widetilde{\Upsilon}_{i}^{10}\neq 0, j=1,\dots, n_{i} \}$, and $A_{i}^{11}=\{ij: Z_{ij}=1, Y_{ij}^{*}=1, \widetilde{\Upsilon}_{i}^{11}\neq \Lambda_{i}^{11}, j=1,\dots, n_{i} \}$.

For type II randomization designs, let $\widetilde{\mathbf{d}}=(\widetilde{d}_{11}, \dots, \widetilde{d}_{S\widetilde{N}_{S}})$ be an optimal solution to (P2). According to Definition 3 in the main text, we have 
\begin{align*}
    &W_{T}^{FP}=\frac{|\{ij: Z_{ij}=1, Y_{ij}^{*}=1, \widetilde{Y}_{ij}=0\}|}{|\{ij: Y_{ij}^{*}\neq \widetilde{Y}_{ij} \}|}=\frac{\sum_{s=1}^{S}\sum_{p=1}^{\widetilde{N}_{s}}\widetilde{d}_{sp}(\Lambda_{[s]}^{11}-\Delta_{sp}^{11})}{\sum_{s=1}^{S}\sum_{p=1}^{\widetilde{N}_{s}}\widetilde{d}_{sp}(\Delta_{sp}^{00}+\Delta_{sp}^{10}+\Lambda_{[s]}^{01}-\Delta_{sp}^{01}+\Lambda_{[s]}^{11}-\Delta_{sp}^{11})}, \\
    &W_{T}^{FN}=\frac{|\{ij: Z_{ij}=1, Y_{ij}^{*}=0, \widetilde{Y}_{ij}=1\}|}{|\{ij: Y_{ij}^{*}\neq \widetilde{Y}_{ij} \}|}=\frac{\sum_{s=1}^{S}\sum_{p=1}^{\widetilde{N}_{s}}\widetilde{d}_{sp}\Delta_{sp}^{10}}{\sum_{s=1}^{S}\sum_{p=1}^{\widetilde{N}_{s}}\widetilde{d}_{sp}(\Delta_{sp}^{00}+\Delta_{sp}^{10}+\Lambda_{[s]}^{01}-\Delta_{sp}^{01}+\Lambda_{[s]}^{11}-\Delta_{sp}^{11})},\\
    &W_{C}^{FP}=\frac{|\{ij: Z_{ij}=0, Y_{ij}^{*}=1, \widetilde{Y}_{ij}=0\}|}{|\{ij: Y_{ij}^{*}\neq \widetilde{Y}_{ij} \}|}=\frac{\sum_{s=1}^{S}\sum_{p=1}^{\widetilde{N}_{s}}\widetilde{d}_{sp} (\Lambda_{[s]}^{01}-\Delta_{sp}^{01})}{\sum_{s=1}^{S}\sum_{p=1}^{\widetilde{N}_{s}}\widetilde{d}_{sp}(\Delta_{sp}^{00}+\Delta_{sp}^{10}+\Lambda_{[s]}^{01}-\Delta_{sp}^{01}+\Lambda_{[s]}^{11}-\Delta_{sp}^{11})},\\
    &W_{C}^{FN}=\frac{|\{ij: Z_{ij}=0, Y_{ij}^{*}=0, \widetilde{Y}_{ij}=1\}|}{|\{ij: Y_{ij}^{*}\neq \widetilde{Y}_{ij} \}|}=\frac{\sum_{s=1}^{S}\sum_{p=1}^{\widetilde{N}_{s}}\widetilde{d}_{sp}\Delta_{sp}^{00}}{\sum_{s=1}^{S}\sum_{p=1}^{\widetilde{N}_{s}}\widetilde{d}_{sp}(\Delta_{sp}^{00}+\Delta_{sp}^{10}+\Lambda_{[s]}^{01}-\Delta_{sp}^{01}+\Lambda_{[s]}^{11}-\Delta_{sp}^{11})}.
\end{align*}
Meanwhile, we can get a collection of sensitive sets
  \begin{align*}
      \bigcup_{g \in G}g\mathcal{S}=\bigcup_{s=1}^{S}\bigcup_{p=1}^{\widetilde{N}_{s}} \big\{ B_{sp}^{00}\cup B_{sp}^{01} \cup B_{sp}^{10} \cup B_{sp}^{11}\big\},
  \end{align*}
  where $B_{sp}^{00}=\{ij: Z_{ij}=0, Y_{ij}^{*}=0, \mathbf{\Lambda}_{i}=\mathbf{\Lambda}_{[s]}, \widetilde{d}_{sp}\neq 0, \Delta_{sp}^{00}\neq 0 \}$, $B_{sp}^{01}=\{ij: Z_{ij}=0, Y_{ij}^{*}=1, \mathbf{\Lambda}_{i}=\mathbf{\Lambda}_{[s]}, \widetilde{d}_{sp}\neq 0, \Delta_{sp}^{01}\neq \Lambda_{[s]}^{01} \}$, $B_{sp}^{10}=\{ij: Z_{ij}=1, Y_{ij}^{*}=0, \mathbf{\Lambda}_{i}=\mathbf{\Lambda}_{[s]}, \widetilde{d}_{sp}\neq 0, \Delta_{sp}^{10}\neq 0 \}$, and $B_{sp}^{11}=\{ij: Z_{ij}=1, Y_{ij}^{*}=1, \mathbf{\Lambda}_{i}=\mathbf{\Lambda}_{[s]}, \widetilde{d}_{sp}\neq 0, \Delta_{sp}^{11}\neq \Lambda_{[s]}^{11} \}$.

\section*{Web Appendix D: Additional Simulation Studies and Related Discussions}

In the main text, we conducted simulation studies on the computational efficiency of the adaptive reformulation strategy for calculating warning accuracy and sensitivity weights proposed in Section 4.2 of the main text with Fisher's sharp null. We also obtained some insights on how warning accuracy and sensitivity weights vary with the effect size of measured outcomes and sample size. In this section, we conduct parallel simulation studies with Neyman's weak null. We investigate both type I and type II randomization designs by considering Simulation Scenario 1, proposed in Section 4.3 of the main text, and Simulation Scenario 3, described as follows. As emphasized in the main text, all the data-generating processes described in the simulation studies in this paper are only for automatically generating data sets for simulations, as our framework works for finite-population data sets and does not depend on specific data-generating models.

\begin{itemize}
     \item \textbf{Simulation Scenario 3 (for Type II randomization designs):} We consider a stratified randomized experiment or a stratified observational study (with most strata being small) with $I=400$ or $2000$ strata. We let $\mathcal{U}(A)$ denote the uniform distribution over the set $A$. In each independent simulation run, for each $i=1,\dots, I$ we randomly draw $m_{i}$ from $\mathcal{U}(\{2, 3\})$ and then randomly draw $n_{i}-m_{i}$ from $\mathcal{U}(\{2, 3\})$. Then we have $E(N)=\{E(m_{i})+E(n_{i}-m_{i})\}\cdot I=5I=2000$ or $10,000$. 
\end{itemize}

In each independent simulation run, after generating $m_{i}$ and $n_{i}-m_{i}$ for each stratum $i$, we follow the same procedure as described in Section 4.3 of the main text to generate the treatment indicators $\mathbf{Z}$ and the measured outcomes $\mathbf{Y}^{*}$ based on the prespecified measured effect size $(p_{0}, p_{1})$. We here consider testing Neyman's weak null $H_{0}^{\text{weak}}$. After conducting 1000 independent simulation runs for each of the 18 different prespecified sets of $(E(N), p_{0}, p_{1})$ under Simulation Scenarios 1 and 3, we give the simulation results of the corresponding average computation time, average warning accuracy and average sensitivity weights in Table~\ref{tab: simulations for weak null} in Web Appendix D. We here report some related details about the specific procedure for obtaining the results in Table~\ref{tab: simulations for weak null}: (i) As mentioned in the main text, conducting a sensitivity analysis or a validation study is typically more meaningful when we detected a treatment effect in a primary analysis (\citealp{rosenbaum2002observational}) based on measured outcomes $\mathbf{Y}^{*}$. Therefore, we here exclude few simulation runs (i.e., generated data sets) in which Neyman's weak null failed to be rejected based on $\mathbf{Y}^{*}$ (20 out of 36,000 runs). (ii) For the remaining 35,980 simulation runs, to prevent our simulation studies from failing to be finished in a tolerable amount of time, we force a simulation run to stop if it runs more than 100 seconds, report the total number of such cases, and exclude such cases when calculating the average computation time, warning accuracy, and sensitivity weights. It turns out that such potentially computationally infeasible cases (computation time more than 100 seconds) are very rare (17 out of 35,980 runs), and in most cases, our framework is computationally efficient with Neyman's weak null. (iii) As in Section 4.3 of the main text, all the computation tasks in this section were also done by the optimization solver \textsf{Gurobi} (version 9.1) (Gurobi Optimization, LLC, 2022) and a laptop computer with a 1.6 GHz Dual-Core Intel Core i5 processor and 4 GB 1600 MHz DDR3 memory. From the simulation results reported in Table~\ref{tab: simulations for weak null}, we can see that the general patterns observed and the insights obtained from Fisher's sharp null case considered in Table 2 of the main text (see the detailed description in Section 4.3 of the main text) also hold for Neyman's weak null case.

\begin{table}[b]
\scriptsize
\centering \caption{Simulations with Neyman's weak null. We report the average computation time (in seconds), warning accuracy $\mathcal{WA}$ and sensitivity weights $(W_{T}^{FP}, W_{T}^{FN}, W_{C}^{FP}, W_{C}^{FN})$ of different sets of $(E(N), p_{0}, p_{1})$ for Simulation Scenarios 1 and 3 (for type I and type II randomization designs respectively).}
\label{tab: simulations for weak null}
\resizebox{\textwidth}{!}{\begin{tabular}{ccccccccccccc}
\hline
\multicolumn{13}{c}{\textbf{Type I Randomization Designs (Simulation Scenario 1)}} \\
\hline
\multirow{2}{*}{$p_{0}=0.3$} & \multicolumn{6}{c}{$E(N)=2000$} & \multicolumn{6}{c}{$E(N)=10,000$}\\
\cmidrule(r){2-7} \cmidrule(r){8-13} 
& Time & $\mathcal{WA}$ & $W_{T}^{FP}$ & $W_{T}^{FN}$ & $W_{C}^{FP}$ & $W_{C}^{FN}$ & Time & $\mathcal{WA}$ & $W_{T}^{FP}$ & $W_{T}^{FN}$ & $W_{C}^{FP}$ & $W_{C}^{FN}$ \\
\hline
$p_{1}=0.4$ & 0.20 s & 0.98 & 0.32 & 0.00 & 0.00 & 0.68 & 5.62 s & 0.98 & 0.35 & 0.00 & 0.00 & 0.65 \\
$p_{1}=0.6$ & 0.26 s & 0.92 & 0.46 & 0.00 & 0.00 & 0.54 & 6.35 s & 0.91 & 0.46 & 0.00 & 0.00 & 0.54 \\
$p_{1}=0.8$ & 0.35 s & 0.83 & 0.54 & 0.00 & 0.00 & 0.46 & 7.59 s & 0.83 & 0.54 & 0.00 & 0.00 & 0.46 \\
\hline
\multirow{2}{*}{$p_{0}=0.6$} & \multicolumn{6}{c}{$E(N)=2000$} & \multicolumn{6}{c}{$E(N)=10,000$}\\
\cmidrule(r){2-7} \cmidrule(r){8-13} 
& Time & $\mathcal{WA}$ & $W_{T}^{FP}$ & $W_{T}^{FN}$ & $W_{C}^{FP}$ & $W_{C}^{FN}$ & Time & $\mathcal{WA}$ & $W_{T}^{FP}$ & $W_{T}^{FN}$ & $W_{C}^{FP}$ & $W_{C}^{FN}$ \\
\hline
$p_{1}=0.7$ & 0.20 s & 0.98 & 0.66 & 0.00 & 0.00 & 0.34 & 5.65 s & 0.98 & 0.65 & 0.00 & 0.00 & 0.35 \\
$p_{1}=0.8$ & 0.25 s & 0.95 & 0.70 & 0.00 & 0.00 & 0.30 & 6.09 s & 0.94 & 0.69 & 0.00 & 0.00  & 0.31 \\
$p_{1}=0.9$ & 0.29 s & 0.91 & 0.74 & 0.00 & 0.00 & 0.26 & 7.15 s & 0.91 & 0.71 & 0.00 & 0.00 & 0.29 \\
\hline
\multirow{2}{*}{$p_{0}=0.9$} & \multicolumn{6}{c}{$E(N)=2000$} & \multicolumn{6}{c}{$E(N)=10,000$}\\
\cmidrule(r){2-7} \cmidrule(r){8-13} 
& Time & $\mathcal{WA}$ & $W_{T}^{FP}$ & $W_{T}^{FN}$ & $W_{C}^{FP}$ & $W_{C}^{FN}$ & Time & $\mathcal{WA}$ & $W_{T}^{FP}$ & $W_{T}^{FN}$ & $W_{C}^{FP}$ & $W_{C}^{FN}$ \\
\hline
$p_{1}=0.2$ & 0.45 s & 0.75 & 0.00 & 0.46 & 0.54 & 0.00 & 7.83 s & 0.74 & 0.00 & 0.46 & 0.54 & 0.00 \\
$p_{1}=0.4$ & 0.36 s & 0.83 & 0.00 & 0.37 & 0.63 & 0.00 & 8.73 s & 0.83 & 0.00 & 0.39 & 0.61 & 0.00 \\
$p_{1}=0.6$ & 0.29 s & 0.91 & 0.00 & 0.26 & 0.74 & 0.00 & 7.34 s & 0.91 & 0.00 & 0.29 & 0.71 & 0.00 \\
\hline
\multicolumn{13}{c}{\textbf{Type II Randomization Designs (Simulation Scenario 3)}} \\
\hline
\multirow{2}{*}{$p_{0}=0.3$} & \multicolumn{6}{c}{$E(N)=2000$} & \multicolumn{6}{c}{$E(N)=10,000$}\\
\cmidrule(r){2-7} \cmidrule(r){8-13} 
& Time & $\mathcal{WA}$ & $W_{T}^{FP}$ & $W_{T}^{FN}$ & $W_{C}^{FP}$ & $W_{C}^{FN}$ & Time & $\mathcal{WA}$ & $W_{T}^{FP}$ & $W_{T}^{FN}$ & $W_{C}^{FP}$ & $W_{C}^{FN}$ \\
\hline
$p_{1}=0.4$ & 0.63 s & 0.98 & 0.25 & 0.00 & 0.00 & 0.75 & 5.12 s & 0.97  & 0.24 & 0.00 & 0.00 & 0.76 \\
$p_{1}=0.6$ & 0.63 s & 0.90 & 0.45 & 0.00 & 0.00 & 0.55 & 5.11 s & 0.89 & 0.46 & 0.00 & 0.00 & 0.54 \\
$p_{1}=0.8$ & 0.52 s & 0.81 & 0.53 & 0.00 &  0.00 & 0.47 & 4.94 s & 0.80 & 0.53 & 0.00 & 0.00 & 0.47 \\
\hline
\multirow{2}{*}{$p_{0}=0.6$} & \multicolumn{6}{c}{$E(N)=2000$} & \multicolumn{6}{c}{$E(N)=10,000$}\\
\cmidrule(r){2-7} \cmidrule(r){8-13} 
& Time & $\mathcal{WA}$ & $W_{T}^{FP}$ & $W_{T}^{FN}$ & $W_{C}^{FP}$ & $W_{C}^{FN}$ & Time & $\mathcal{WA}$ & $W_{T}^{FP}$ & $W_{T}^{FN}$ & $W_{C}^{FP}$ & $W_{C}^{FN}$ \\
\hline
$p_{1}=0.7$ & 0.62 s & 0.98 & 0.74 & 0.00 & 0.00 & 0.26 & 5.14 s & 0.97 & 0.75 & 0.00 & 0.00 & 0.25 \\
$p_{1}=0.8$ & 0.55 s & 0.94 & 0.68 & 0.00 & 0.00 & 0.32 & 4.95 s & 0.93 & 0.67 & 0.00 & 0.00 & 0.33 \\
$p_{1}=0.9$ & 0.43 s & 0.89 & 0.71 & 0.00 & 0.00 & 0.29 & 4.82 s & 0.89 & 0.70 & 0.00 & 0.00 & 0.30 \\
\hline
\multirow{2}{*}{$p_{0}=0.9$} & \multicolumn{6}{c}{$E(N)=2000$} & \multicolumn{6}{c}{$E(N)=10,000$}\\
\cmidrule(r){2-7} \cmidrule(r){8-13} 
& Time & $\mathcal{WA}$ & $W_{T}^{FP}$ & $W_{T}^{FN}$ & $W_{C}^{FP}$ & $W_{C}^{FN}$ & Time & $\mathcal{WA}$ & $W_{T}^{FP}$ & $W_{T}^{FN}$ & $W_{C}^{FP}$ & $W_{C}^{FN}$ \\
\hline
$p_{1}=0.2$ & 0.56 s & 0.71 & 0.00 & 0.46 & 0.54 & 0.00 & 5.03 s & 0.70 & 0.00 & 0.45 & 0.55 & 0.00 \\
$p_{1}=0.4$ & 0.44 s & 0.81 & 0.00 & 0.40 & 0.60 & 0.00 & 4.84 s & 0.80 & 0.00 & 0.39 & 0.61 & 0.00 \\
$p_{1}=0.6$ & 0.48 s & 0.89 & 0.00 & 0.29 & 0.71 & 0.00 & 4.85 s & 0.89 & 0.00 & 0.29 & 0.71 & 0.00 \\
\hline
\end{tabular}}
\end{table}

\section*{Web Appendix E: Additional Clarifications and Discussions}

Transparently and unambiguously calculated from the dataset, the outputs of our sensitivity analysis framework (i.e., the warning accuracy and sensitivity weights) help suggest whether a randomization test's result is sensitive to outcome misclassification (in the worst-case sense) and whether it is especially sensitive to a certain type of outcome misclassification (in the sense of sensitivity to adversarial attacks of outcome misclassification). Although our sensitivity analysis's outputs are objective, and there is no ``right or wrong" for them (as long as they can be correctly understood and exactly calculated), researchers should be careful to avoid misinterpreting them and be aware that such sensitivity analysis information should be combined with domain knowledge and/or validation studies to better inform the final causal decision-making.

Specifically, recall that the warning accuracy only tells us the minimal number of outcome misclassification cases needed to overturn a causal conclusion reported by a randomization test. It does not tell us whether there exist many outcome misclassification cases in the dataset, as such information is beyond the scope of sensitivity analysis and involves the actual implementation of a scientific investigation or validation study. When combining the warning accuracy and subject knowledge, there are four main cases:
\begin{itemize}
    \item Case 1.1: The sensitivity analysis suggests that the original randomization test's result is sensitive to outcome misclassification (i.e., exhibits high warning accuracy). Meanwhile, domain knowledge or validation studies suggest that we may expect many outcome misclassification cases in the dataset. 
    
    \item Case 1.2: The sensitivity analysis suggests that the original randomization test's result is sensitive to outcome misclassification (i.e., exhibits high warning accuracy). Meanwhile, domain knowledge or validation studies suggest that outcome misclassification should rarely exist in the dataset. 
    
    \item Case 1.3: The sensitivity analysis suggests that the original randomization test's result is insensitive (robust) to outcome misclassification (i.e., exhibits low warning accuracy). Meanwhile, domain knowledge or validation studies suggest that many outcome misclassification cases may exist in the dataset.  
    
    \item Case 1.4: The sensitivity analysis suggests that the original randomization test's result is insensitive (robust) to outcome misclassification (i.e., exhibits low warning accuracy). Meanwhile, domain knowledge or validation studies suggest that outcome misclassification should rarely exist in the dataset.  
\end{itemize}

Also, recall that another quantity reported in our framework, the sensitivity weights, only suggests whether a randomization test's result is especially sensitive to one of the following four types of outcome misclassification: false positives among the treated group, false negatives among the treated group, false positives among the control group, and false negatives among the control group. The sensitivity weights do not tell us which of the aforementioned four types of outcome misclassification is expected to occur more in the dataset. This question involves conducting a scientific investigation or validation study and cannot be answered by data analysis alone. When combining the sensitivity weights and subject knowledge, there are three main cases:
\begin{itemize}
    \item Case 2.1: The sensitivity weights suggest that the original randomization test's result is especially sensitive to a certain type of outcome misclassification. Meanwhile, domain knowledge or validation studies suggest that that type of outcome misclassification is expected to occur more frequently in the dataset than other types of outcome misclassification. 
    
    \item Case 2.2: The sensitivity weights suggest that the original randomization test's result is especially sensitive to a certain type of outcome misclassification. Meanwhile, domain knowledge or validation studies suggest that that type of outcome misclassification is \textit{not} expected to occur more frequently than other types of outcome misclassification. 
    
    \item Case 2.3: From the sensitivity weights, we do not have evidence that the randomization test's result is especially sensitive to a certain type of outcome misclassification.
    
\end{itemize}

The causal evidence from the original randomization test is weakened in Cases 1.1 and 2.1 (the investigation for the observation that ``finasteride increases the risk of high-grade prostate cancer" in the PCPT study presented in Section 5 belongs to these two cases) and strengthened in Case 1.4. In Cases 1.2, 1.3, 2.2, and 2.3, there is no immediate implication about whether the evidence from the original randomization test should be weakened or strengthened, which provides useful suggestions about the necessity of further validation studies and investigations on potential bias from outcome misclassification. In summary, in all of the above cases, our framework can help researchers get a more comprehensive view of the result from the original randomization test. 

\endgroup

\end{document}